%% file: sn-article.tex
\documentclass[pdflatex,sn-nature]{sn-jnl}

\usepackage{graphicx}%
\usepackage{multirow}%
\usepackage{amsmath,amssymb,amsfonts}%
\usepackage{amsthm}%
\usepackage[title]{appendix}%
\usepackage{xcolor}%
\usepackage{textcomp}%
\usepackage{manyfoot}%
\usepackage{booktabs}%
\usepackage{algorithm}%
\usepackage{algorithmicx}%
\usepackage{algpseudocode}%
\usepackage{listings}%

\usepackage{tabularx}
\usepackage{array}
\usepackage{adjustbox}
\usepackage{longtable}

\usepackage{graphicx}
\usepackage{makecell}
\usepackage{booktabs}

\setcitestyle{super,open={},close={},sort&compress}

\theoremstyle{thmstyleone}%
\theoremstyle{thmstyletwo}%

\theoremstyle{thmstylethree}%

\hypersetup{hypertexnames=false}

\newcommand{\revise}[1]{#1}
\newcommand{\todayrevise}[1]{#1}
\newcommand{\latestrevise}[1]{#1}

\begin{document}
\title[Informal learning emerges in human--LLM interaction]{\latestrevise{Informal learning emerges in everyday human--LLM interaction}}


\Newlabel{zixin-intern}{1†}
\author[zixin-intern]{\fnm{Zixin} \sur{Chen}}\email{zchendf@connect.ust.hk}

\author*[2]{\fnm{Haotian} \sur{Li}}\email{haotian.li@microsoft.com}

\author[3]{\fnm{Ziang} \sur{Xiao}}\email{ziang.xiao@jhu.edu}
\author[1]{\fnm{Huamin} \sur{Qu}}\email{huamin@cse.ust.hk}
\author[2]{\fnm{Xing} \sur{Xie}}\email{xxie2023@outlook.com}

\affil[1]{\orgname{The Hong Kong University of Science and Technology}, \orgaddress{\city{Hong Kong}, \country{China}}}

\affil[2]{\orgname{Microsoft Research Asia}, \orgaddress{\city{Beijing}, \country{China}}}

\affil[3]{\orgname{Johns Hopkins University}, \orgaddress{\city{Baltimore}, \postcode{21218}, \state{Maryland}, \country{United States}}}

\makeatletter\gdef\@presentaddresstxt{}\makeatother
\presentaddress{\footnotesize $^{\dagger}$This work was done during Z.C.'s internship at Microsoft Research Asia.}

\input{sections/00-abstract}




\maketitle

\input{sections/01-introduction}

\input{sections/02-results_updated}
\input{sections/03-discussion}
\newpage
\input{sections/04-methods_reorganized}

\backmatter

\input{sections/06-declarations}
\newpage

\bibliography{sn-bibliography}

\input{sections/05_supplementary_revised}

\end{document}

%% file: sections/00-abstract.tex
\abstract{

\latestrevise{As LLMs become increasingly capable of completing tasks for users, a central concern is that everyday AI use may become primarily cognitive offloading, eroding the opportunities through which people develop their own capabilities. We analyse large-scale human--LLM conversations to ask whether informal learning behaviors also emerge in this setting: whether users engage in exchanges in ways that preserve opportunities to learn.} Across 128,569 naturalistic conversations, we translated learning-science constructs into turn-level behavioural signatures. Cognitive engagement, users' cognitive effort as reflected in the exchange, appeared in 31.9\% of 491,685 user turns, whereas constructive engagement, the deepest observable form of learning-oriented engagement, appeared in 4.9\%, showing that deeper sense-making was recurrent but selective. Our study further identifies factors associated with these forms of engagement. Scaffolded assistant support consistently marked richer constructive participation, with associations varying by user framing, task ecology, support form, timing and prior user state. \latestrevise{Together, these findings show that everyday human--LLM interaction is not only answer delivery or cognitive offloading; it also contains measurable, selective and conditionally organized behavioural signatures of informal learning.} They shift AI evaluation from answer-delivery efficiency toward the preservation of cognitive opportunities for users to reason, test ideas and construct understanding in the course of everyday problem-solving.

}

%% file: sections/01-introduction.tex
\section{Introduction}\label{sec1}

Large language models (LLMs) are becoming embedded in the cognitive work of everyday life: people increasingly turn to them to write, code, search and reason through unfamiliar problems~\cite{chatterji2025people,noy2023experimental,brynjolfsson2025generative,costaGomes2026healthqueries}. Much of human competence, however, is acquired precisely by doing such work: reasoning through a problem, attempting a solution and reworking it until it holds~\cite{anzai1979theory,kolb2014experiential}. A growing body of evidence suggests that delegating this effort to machines can weaken the very capabilities that the work would otherwise develop~\cite{risko2016cognitive,gerlich2025ai}. Whether assistance displaces users' cognitive effort or becomes an occasion for further thought depends not only on what the technology does, but also on how people engage with it: they may accept an answer and move on, or use it to test their own ideas, seek explanations and probe a question more deeply~\cite{barke2023grounded,kobiella2024if}. \revise{Prior work characterizes informal learning as learning that occurs outside formally structured curricula, instruction or assessment, often as part of everyday activity and whether or not learning is the person's primary objective~\cite{eraut2004informal,marsick2015informal,schugurensky2000forms}. In human--LLM interaction, such learning may arise as users confront uncertainty, seek explanations, test ideas and revise their understanding. Everyday human--LLM interaction is therefore a vast, largely undesigned setting for informal learning. As AI assumes more of the cognitive work through which competence is built, the stakes extend beyond immediate task performance to whether users continue to exercise and develop their own capabilities.}

Yet informal learning in general-purpose human--LLM interaction remains largely unexamined. Existing research on LLMs and learning has focused instead on settings in which learning support is deliberately designed, such as automated feedback, question generation, retrieval practice and tutoring systems~\cite{kasneci2023chatgpt,meyer2024using,arif2024generation,jin2024teach,lim2024identify,yan2024promises,futterer2026learningcoach}. These studies show that LLMs can enact instructional strategies when learning is an explicit design target~\cite{terzimehic2026conversational,vanlehn2011relative}; they tell us less about general-purpose use, in which users typically seek progress on a task rather than instruction. Everyday use is also harder to study: it follows no curriculum and yields no direct measure of knowledge change, making learning easy to posit but difficult to observe. Learning science offers a process-level lens: deeper forms of overt engagement are associated with stronger learning outcomes, allowing learning-oriented engagement to serve as an observable process indicator of informal learning behaviors in user-generated interactions whose goals and endpoints are not specified by a curriculum or study protocol~\cite{fredricks2004school,chi2014icap,koedinger2007exploring}. We therefore ask whether learning-oriented engagement behaviors exist in everyday human--LLM interaction, where its most constructive forms arise, and what factors may shape these behaviors.

We analyse 128,569 naturalistic conversations from three public large-scale conversational corpora: WildChat, LMSYS Chat and ShareChat~\cite{zhao2024wildchat,zheng2023lmsyschat1m,yan2025sharechat}. Across these corpora, we focus on coding and writing as two common task ecologies in which LLMs are used to produce, revise and troubleshoot knowledge work~\cite{chatterji2025people, handa2025economic, barke2023grounded}, yielding 491,685 user turns and 489,785 assistant turns. Our analyses move from engagement prevalence, to the contextual organization of constructive engagement, to how assistant scaffolding is associated with user participation at conversation-level, support-form and adjacent-turn scales. To make these questions measurable in naturalistic logs, we translate theories of engagement and scaffolding into role-specific, turn-level behavioural signatures. \revise{User turns are coded by the depth of users' cognitive effort as reflected in the exchange: \textit{passive} receipt, \textit{active} use or follow-up, and \textit{constructive} engagement, where users elaborate, test, revise or extend ideas in the exchange~\cite{fredricks2004school,chi2014icap}. Constructive engagement is therefore the clearest observable signature of informal learning in these logs: it marks moments when task progress becomes explicit knowledge work rather than simple answer use. Assistant turns are coded by whether they primarily deliver the requested content or provide scaffolded support: contingent assistance that diagnoses difficulties, decomposes tasks, explains rationales, gives feedback or hints, or prompts reflection while leaving the user to continue the work~\cite{wood1976role,van2010scaffolding,winkler2020sara,quintana2018scaffolding}. Together, these signatures allow us to examine where users remain actively involved in reasoning through task material, and how constructive participation is organized by task context, assistant support and local turn ordering.}




\begin{figure*}[t]
    \centering
    \includegraphics[width=\textwidth]{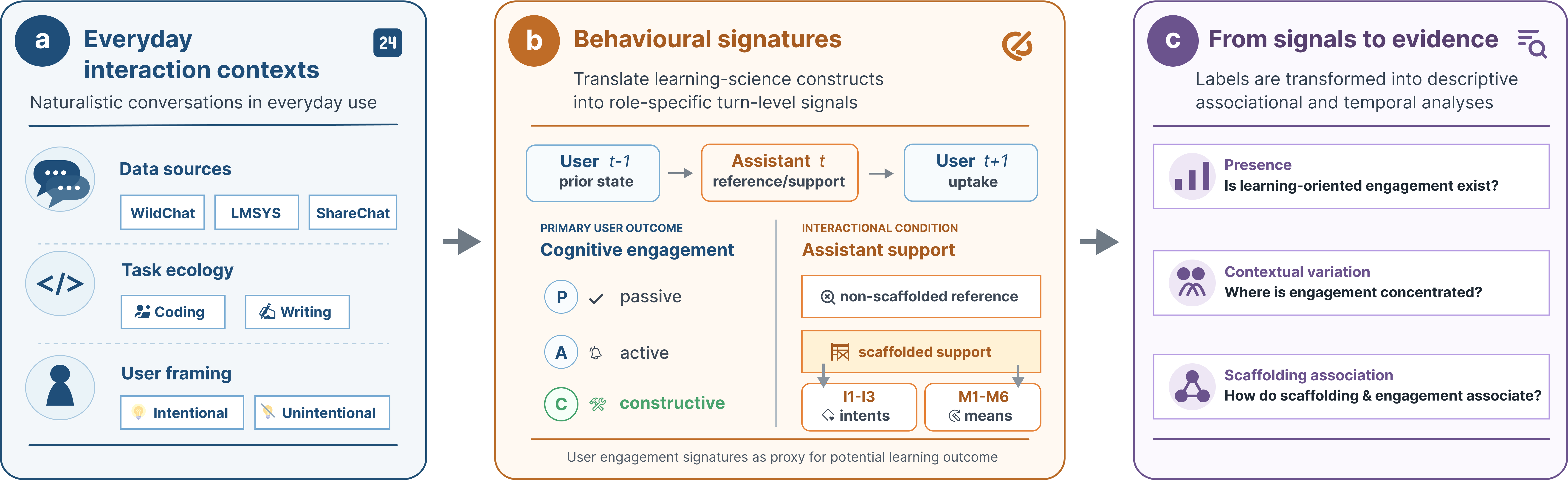}
    \caption{
\textbf{Analytic framework for studying informal learning in everyday human--LLM interaction.}
\textbf{a,} Public conversations are situated by data source, task ecology and user framing.
\textbf{b,} Engagement and scaffolding constructs are translated into role-specific turn-level labels for user engagement and assistant support\revise{; passive, active and constructive labels form depth levels within cognitive engagement, while user framing is a conversation-level label shown in panel a. Assistant support is further characterized by support intent (I1--I3: metacognitive, cognitive and affective) and support form (M1--M6: feedback, hinting, instructing, explaining, modelling and questioning); Supplementary Table~\ref{tab:label_examples} provides plain-language examples of each label}.
\textbf{c,} These labels support analyses of engagement presence, contextual organization and support--engagement coupling at conversation-level, support-form and adjacent-turn scales\revise{; adjacent-turn analyses examine the immediately preceding user state, the assistant response and the next user turn}.
Labels indicate observable learning-oriented participation, not durable learning outcomes or causal instructional effects.
}
    \label{fig:conceptual_framework}
\end{figure*}

Across the three corpora, everyday human--LLM interaction emerged as a genuine but selective and conditional site of learning. First, learning-oriented engagement was recurrent in everyday human--LLM interaction, but its deepest constructive form was selective: cognitive engagement appeared in nearly one third of user turns, whereas constructive engagement appeared in about five percent. Second, constructive engagement was ecologically organized: it was more common under explicit learning-oriented user framing, varied across task ecologies and became most visible in sustained exchanges where users returned to, interrogated and reshaped task materials. Third, assistant scaffolding marked richer constructive participation, but not as a uniform exposure. Its association depended on how support was positioned within the exchange: feedback-like and explanatory forms aligned most clearly with constructive engagement at the conversation level, while adjacent-turn analyses showed that support--engagement coupling in local sequences was shaped by prior user state and support form. Together, these findings locate everyday human--LLM interaction in a middle space between ordinary tool use and designed instruction. \latestrevise{They also sharpen the broader stakes of AI-mediated work beyond simply cognitive offloading or answer consumption: everyday AI use is a setting in which informal learning can emerge when users remain engaged in questioning, testing and revising ideas.} 
They suggest that the educational value of future AI assistants lies not in making every interaction tutor-like, but in calibrating when to answer, explain, give feedback or step back, preserving the cognitive agency through which people question, test, revise and build judgment while solving real problems.


%% file: sections/02-results_updated.tex
\section{Results}\label{sec:results}


The results follow the analytic sequence in Fig.~\ref{fig:conceptual_framework}: engagement prevalence, contextual organization and support--engagement coupling. We first characterize cognitive and constructive engagement and examine how these forms of engagement vary across interaction contexts, \revise{then analyse the association between assistant support and constructive participation across conversations and within local user--assistant--user sequences.}



\subsection{Everyday human--LLM conversations contain learning-oriented engagement}\label{subsec:2.1}

Learning-oriented engagement was visible across all six task settings. Across 491,685 user turns, 31.9\% showed cognitive engagement and 4.9\% reached constructive engagement, \revise{a subset of cognitive engagement and} the deepest observable user-side signal in the annotation scheme (Table~\ref{tab:dataset_overview}). Thus, everyday human--LLM conversations often contained observable user participation in the exchange, but deeper constructive sense-making remained selective rather than routine.

\begin{figure*}[t]
    \centering
    \includegraphics[width=\textwidth]{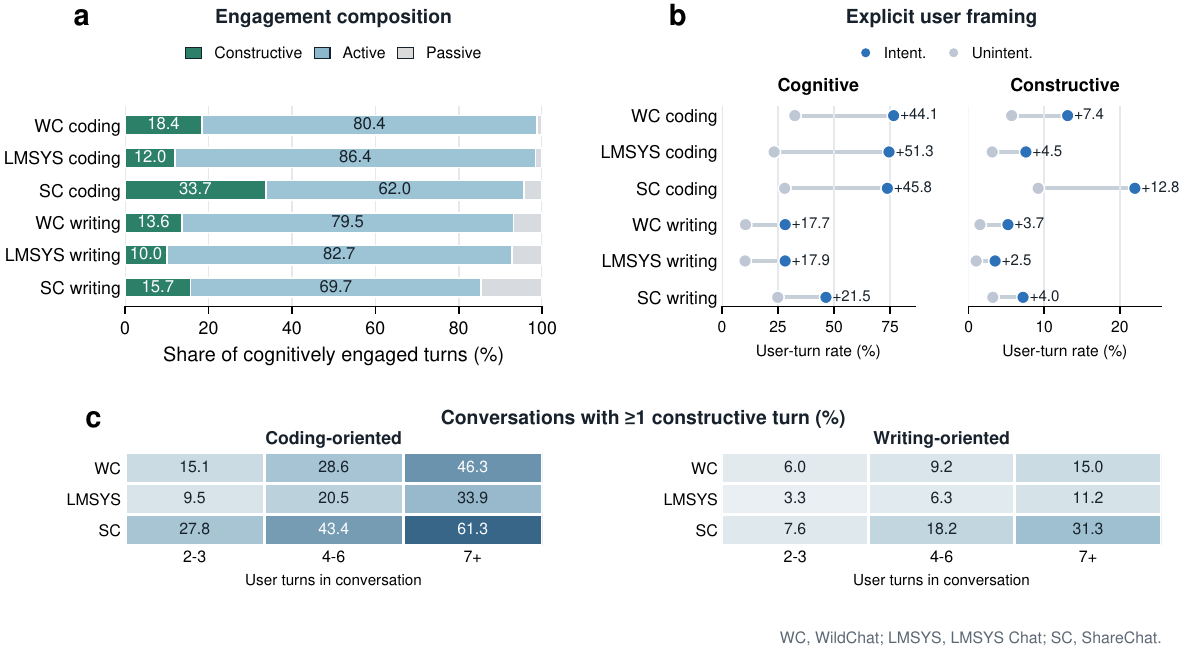}
    \caption{
    \textbf{Learning-oriented engagement varies by engagement form, user framing and conversation length.}
\textbf{a,} Composition of cognitively engaged user turns by passive, active and constructive levels across the six task settings.
\textbf{b,} Cognitive and constructive user-turn rates in conversations with versus without explicit learning-oriented framing; annotations show the framed-minus-unframed difference in percentage points.
\textbf{c,} Share of conversations containing at least one constructive user turn by conversation-length bucket.
Panel a decomposes cognitively engaged user turns\revise{, where constructive engagement is nested within cognitive engagement}; panel b reports percentages of user turns; panel c reports conversation-level shares. WC, WildChat; LMSYS, LMSYS Chat; SC, ShareChat.
}
    \label{fig:engagement_ecology}
\end{figure*}

Figure~\ref{fig:engagement_ecology}a shows how cognitively engaged turns were composed. Active engagement was the dominant form in every setting, accounting for 62.0--86.4\% of cognitively engaged turns, whereas constructive engagement accounted for 10.0--33.7\% and passive receipt for 1.2--14.6\%. This composition separates broad engagement from deeper sense-making. Cognitive engagement shows that everyday LLM use often involves users working within the exchange; active engagement captures task-relevant use, application or follow-up; and constructive engagement identifies the cases in which users elaborate, test, revise or extend ideas. The remaining analyses therefore focus on where constructive engagement appears and how it is associated with user framing, task ecology, interaction depth and assistant support.

\input{tables/table3}

\subsection{Constructive engagement is patterned by user framing, task ecology and sustained exchange}\label{subsec:2.2}

Having established that everyday human--LLM conversations contain learning-oriented engagement, we next examined where its constructive form became visible. Explicit learning-oriented framing provided the first contrast (Fig.~\ref{fig:engagement_ecology}b). In every setting, conversations framed around learning had higher cognitive and constructive user-turn rates than conversations without such framing. The cognitive-engagement lift ranged from +17.7 to +51.3 percentage points across settings, and the constructive-engagement lift ranged from +2.5 to +12.8 points. Yet constructive participation was also present in conversations without explicit learning-oriented framing. User framing therefore amplified constructive engagement, but did not define it: learning-relevant participation also emerged in task-oriented exchanges where learning was incidental to solving the problem at hand.

Task ecology provided a second source of variation. Across all three corpora, coding-oriented conversations showed higher cognitive and constructive engagement than writing-oriented conversations. Cognitive engagement ranged from 45.00--50.21\% in coding conversations, compared with 15.30--33.05\% in writing conversations; constructive engagement ranged from 5.44--15.17\% in coding and 1.53--5.18\% in writing. A conversation-level context model showed the same direction after including user framing, conversation-length bucket and dataset fixed effects, with coding task ecology positively associated with the presence of at least one constructive user turn (OR=1.89, 95\% CI 1.71--2.08, p$<.001$; Supplementary Table~\ref{tab:constructive_context_logit}).

The coding--writing contrast points to task affordances that organize baseline opportunities for visible constructive participation. Coding tasks often externalize uncertainty through errors, constraints and executable tests, creating occasions for users to diagnose failures, refine assumptions, compare alternatives and test candidate solutions. Writing tasks also involve substantial revision and evaluation, but the interaction can often proceed through accepting, adapting or requesting a draft revision without requiring the user to articulate the rationale for the change~\cite{keuning2018systematic,flower1981cognitive,sommers1980revision}. The observed task difference is therefore consistent with the ecological interpretation: constructive engagement is more likely to appear when the task makes uncertainty, evaluation and revision explicit in the exchange.

Finally, sustained exchange provided a third context in which constructive participation became visible (Fig.~\ref{fig:engagement_ecology}c). Across all six settings, longer conversations were more likely to contain at least one constructive user turn. The association was also evident in a grouped-binomial rate model using total user turns as the denominator: relative to conversations with 2--3 user turns, the per-turn odds of constructive engagement were 13\% higher in conversations with 4--6 user turns (OR=1.13) and 6\% higher in conversations with 7 or more user turns (OR=1.06; Supplementary Table~\ref{tab:constructive_context_logit}). Together, these results show that sustained exchanges were associated not only with a greater likelihood of constructive participation, but also with a modestly higher constructive-turn rate within the exchange. Longer exchanges give users room to revisit the assistant's response, introduce new constraints, test implications, or request revisions.

To summarize, the findings show that constructive engagement was ecologically organized rather than evenly distributed across everyday use. It was amplified by explicit learning-oriented framing, patterned by task affordances that create occasions for testing and revision, and most visible in sustained exchanges.

\subsection{Scaffolded support marks richer constructive participation}
\label{subsec:2.3}

\begin{figure*}[t]
  \centering
  \includegraphics[width=\linewidth]{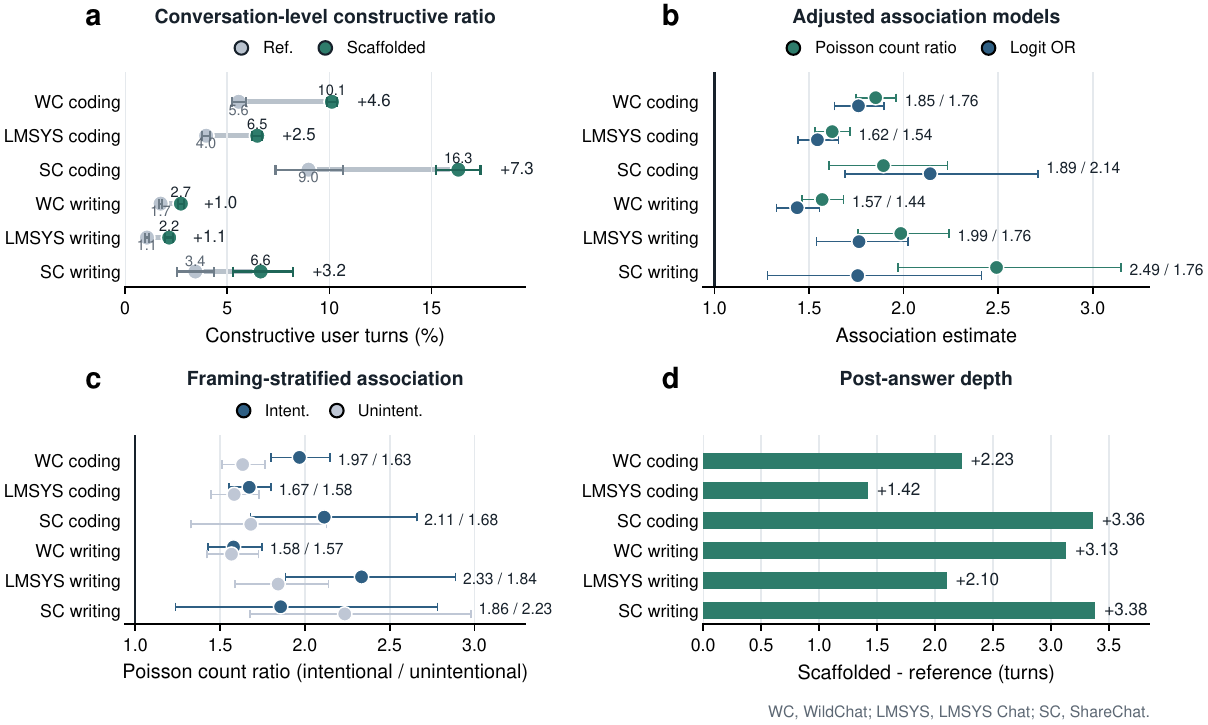}
    \caption{
\textbf{Scaffolded support marks richer constructive participation across task settings.}
\textbf{a,} Turn-weighted constructive user-turn ratios in conversations containing scaffolded support versus reference conversations without scaffolded support.
\textbf{b,} Covariate-adjusted within-setting associations between scaffolded-support presence and constructive participation, reported as Poisson count ratios and logistic odds ratios.
\textbf{c,} Poisson count ratios for scaffolded-support presence stratified by explicit versus non-explicit learning-oriented framing.
\textbf{d,} Difference in the number of turns after the first assistant response between scaffolded and reference conversations.
Numbers indicate scaffolded-minus-reference differences, except panels b and c, which report model estimates. Error bars in panel a show conversation-level bootstrap 95\% confidence intervals for the displayed group ratios; panels b and c show model-based 95\% confidence intervals. Bootstrap confidence intervals for the descriptive contrasts in panels a and d are reported in Supplementary Table~\ref{tab:key_contrast_significance}. WC, WildChat; LMSYS, LMSYS Chat; SC, ShareChat.
}

  \label{fig:s2_conversation_association}
\end{figure*}

The previous section focuses on how constructive participation is associated with users' conditions.
Now we further examine how the assistant's response will affect users' constructive engagement, through the lens of scaffolding.
Scaffolding is a common educational strategy for increasing students' opportunities to participate through approaches such as raising questions or providing hints rather than definitive answers~\cite{wood1976role,van2010scaffolding,winkler2020sara,quintana2018scaffolding}.
In human--LLM conversations, the lens distinguishes LLM responses that simply return a deliverable from those that encourage users' participation, surfacing a critical difference in LLMs' behavior that may affect users' engagement.


\revise{Scaffolded support appeared in 31.7\% of assistant turns (Table~\ref{tab:dataset_overview}), making it a common interactional condition rather than a rare pedagogical exception.} At the conversation level, conversations containing at least one scaffolded assistant turn had higher turn-weighted constructive user-turn ratios than reference conversations without scaffolded support in all six task settings (Fig.~\ref{fig:s2_conversation_association}a). The constructive-ratio differences ranged from +0.99 to +7.34 percentage points across WildChat, LMSYS Chat and ShareChat, and all six contrasts were distinguished from zero in conversation-level bootstrap tests ($p<.001$; Supplementary Table~\ref{tab:key_contrast_significance}). Scaffolded conversations also showed more turns after the first assistant response in every setting (+1.42 to +3.38 turns; Fig.~\ref{fig:s2_conversation_association}d), indicating that scaffolded support marked exchanges with more sustained participation. Together, these descriptive contrasts show that scaffolded support co-occurred with richer constructive participation across corpora and task settings.

The association remained positive after accounting for observed user framing and interactional context in covariate-adjusted models. In setting-specific models, scaffolded-support presence was associated with higher constructive-turn counts, with Poisson count ratios ranging from 1.569 to 2.491, and with higher odds that a conversation contained at least one constructive user turn, with logistic odds ratios ranging from 1.437 to 2.140 (Fig.~\ref{fig:s2_conversation_association}b). The same positive direction appeared within both explicitly learning-oriented and non-explicitly framed conversations (Fig.~\ref{fig:s2_conversation_association}c), indicating that the support--engagement association was not limited to interactions users began with explicit learning-oriented framing. A rate sensitivity treating user turns as the exposure preserved positive scaffolded-support associations in all six settings (rate ratios 1.356--1.703; all $p<.001$; Supplementary Table~\ref{tab:fig3_offset_rate_sensitivity}). Across these checks, scaffolded support remained a consistent conversation-level factor of richer constructive participation.

\subsection{Support form differentiates the scaffolding--engagement association}\label{subsec:support_form}\label{subsec:2.4}

\begin{figure*}[h!]
  \centering
  \includegraphics[width=\linewidth]{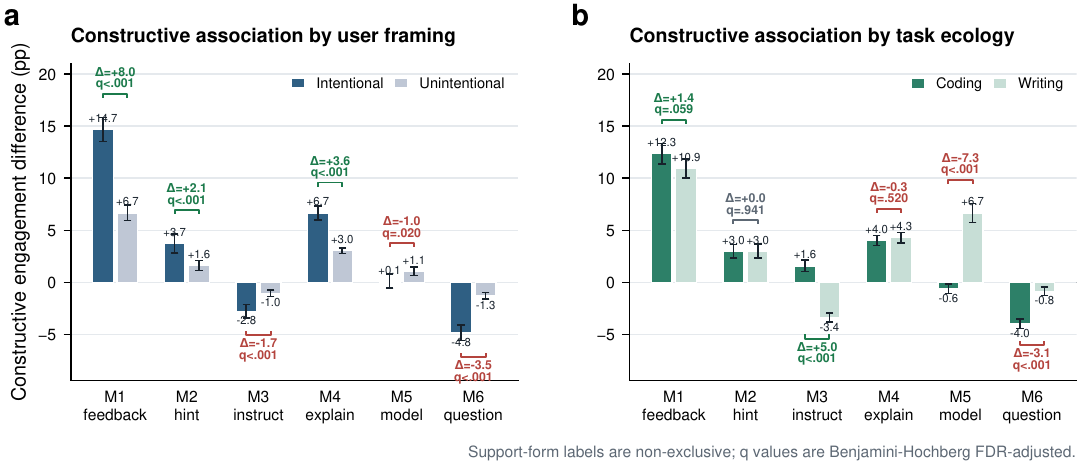}
    \caption{
    \textbf{Support forms differentiate scaffolded interaction.}
    \textbf{a,} Constructive-engagement differences associated with each support form among scaffolded conversations, stratified by intentional versus unintentional user framing.
    \textbf{b,} The same support-form contrasts stratified by coding- versus writing-oriented task ecology.
    Bars are turn-weighted constructive user-turn ratio differences, comparing scaffolded conversations containing a support form with scaffolded conversations without that form; error bars show 95\% confidence intervals, and bracket annotations show between-stratum differences and Benjamini--Hochberg FDR-adjusted q values within each panel family. Support-form labels correspond to M1--M6 and are not mutually exclusive.
    }
  \label{fig:support_form_supply}
\end{figure*}

The conversation-level analyses established scaffolded support as a consistent factor of richer constructive participation. We next asked which features of scaffolded support carried this signal. For each scaffolded assistant turn, the annotation scheme recorded two parallel descriptors: a support-intent label indicating whether the turn primarily targeted metacognitive, cognitive or affective support (I1--I3), and one or more support-form labels indicating how the support was delivered, including feedback, hinting, instructing, explaining, modelling or questioning (M1--M6). This two-axis scheme follows scaffolding accounts that distinguish the purpose of assistance from the means through which assistance is provided~\cite{wood1976role,van2010scaffolding,quintana2018scaffolding,winkler2020sara}. \revise{Support intent offered limited contrast because scaffolded turns were predominantly cognitive in purpose across the three corpora, whereas support-form labels were non-exclusive (Supplementary Fig.~\ref{fig:supp_support_supply}).} We therefore focused on the form axis, comparing scaffolded conversations containing each support form with scaffolded conversations without that form and reporting turn-weighted constructive user-turn ratio differences (Fig.~\ref{fig:support_form_supply}a,b).

At the conversation level, support form sharply differentiated where the scaffolded-support association was concentrated. At this conversation-level scale, feedback-like support (M1) showed the largest constructive-engagement contrast in both explicitly framed and non-explicitly framed conversations, with the association strongest under explicit learning-oriented framing (+14.7 versus +6.7 percentage points; framing difference, +8.0 points; Fig.~\ref{fig:support_form_supply}a). Explaining (M4) showed the same direction (+6.7 versus +3.0 points; framing difference, +3.6 points), and hinting (M2) was positive but smaller. Instructing (M3) and questioning (M6), by contrast, were lower or negative, especially in explicitly framed conversations. Task-stratified estimates gave a complementary view (Fig.~\ref{fig:support_form_supply}b): feedback-like support (M1) and explaining (M4) were positive in both coding- and writing-oriented task ecologies, whereas instructing (M3), modelling (M5) and questioning (M6) varied more strongly by task.

The overall form-level pattern is distinctive because these support moves occurred within general-purpose task assistance rather than within a designed lesson. In this setting, support is interleaved with task completion, so its significance lies partly in how it distributes cognitive responsibility. The positive contrasts for feedback-like support, explanation and, more modestly, hinting are consistent with forms that leave part of the epistemic work available to the user. Feedback turns a user's prior attempt into an object for evaluation, repair or extension~\cite{hattie2007power,shute2008focus,nicol2006formative}; explanation exposes mechanisms or rationales for inspection, testing and adaptation~\cite{chi1989self,renkl1997learning}; and hints leave part of the next inference or task step to the user. Instructions, models and questions can also scaffold progress by specifying an action, demonstrating a pattern or eliciting a response, but they do not necessarily require explicit user evaluation or revision. Their weaker or more variable alignment is consistent with assistance-dilemma accounts in which support can advance task performance while reducing opportunities for user-generated reasoning~\cite{koedinger2007exploring,aleven2003help,reiser2004scaffolding}.

User framing further differentiated the form-level pattern. Under explicit learning-oriented framing, the contrasts among support forms became more pronounced: the positive associations for feedback-like support, explaining and hinting were larger, while instructing and questioning diverged further in the opposite direction; modelling changed little. This wider separation localizes the framing association to the fit between the user's orientation and the form of support provided. When understanding is salient as an interactional goal, feedback and explanation make a user's attempt or the grounds of an answer available for evaluation, repair and extension, whereas directive and questioning moves primarily organize the next action or issue to be addressed. Under non-explicit framing, support was more closely embedded in accomplishing the immediate task, and the distinctions among forms were correspondingly less pronounced.

Task ecology revealed a more specific form--task fit, centred on the crossover between instructing and modelling. Instructing was positively associated with constructive engagement in coding but negatively associated in writing, whereas modelling was approximately null in coding but strongly positive in writing. This crossover may reflect the different feedback structures of the two task ecologies. In coding, carrying out a procedural instruction yields an externally checkable output or error that users can inspect, diagnose and bring back into the exchange. Writing revisions, by contrast, are evaluated through interpretive comparison rather than an equally determinate execution signal. In this ecology, modelling supplies a reusable pattern or worked example intended for adaptation, externalizing linguistic and rhetorical choices that users can compare, select, revise or recombine. The crossover therefore points to two task-specific routes to constructive follow-up: testing an instructed procedure in coding and adapting a modelled exemplar in writing. Thus, in everyday human--LLM interaction, support form is informative not only as a pedagogical category but also as an interactional position. It indicates whether the assistant's contribution becomes material for the user's next cognitive move or advances the task without requiring substantial further reasoning from the user.


Together, the results suggest that scaffolded support marked richer constructive participation, and the conversation-level association was concentrated around particular support forms within task-specific scaffolding repertoires.

\begin{figure}[!htbp]
  \centering

  \includegraphics[width=\textwidth]{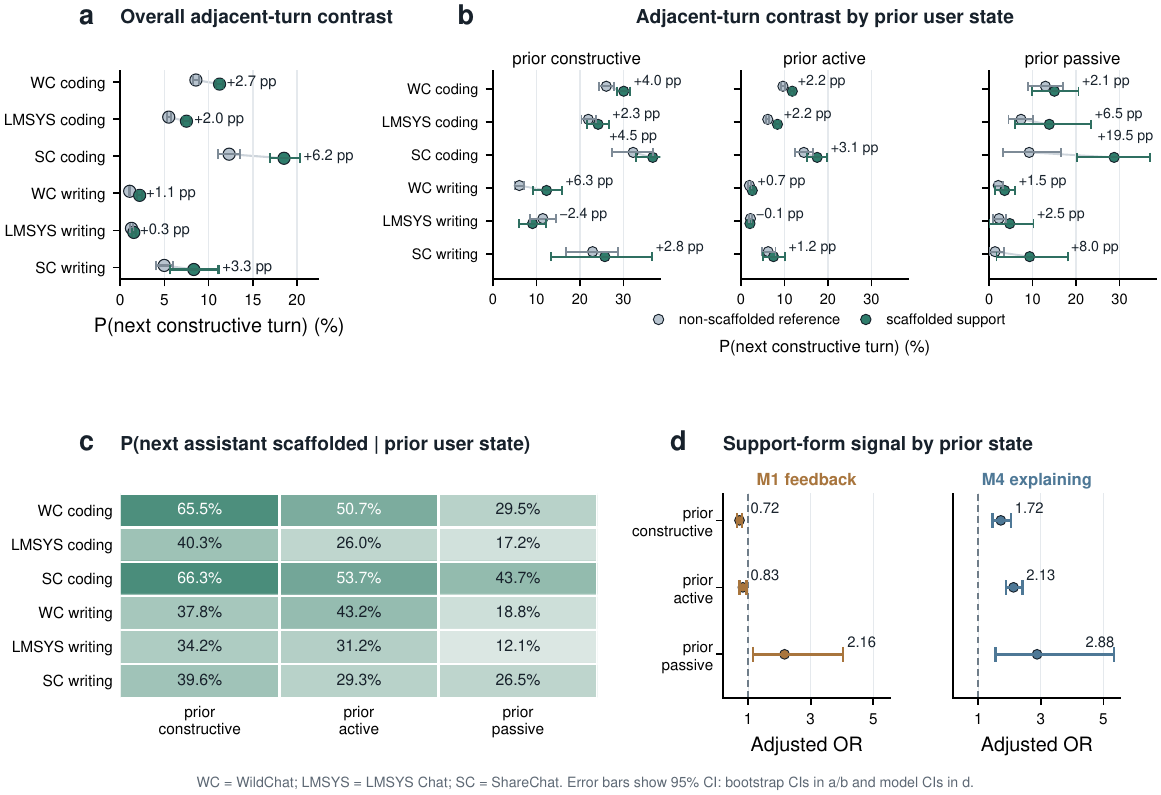}
    \caption{
\textbf{Support--engagement coupling is state- and form-dependent across task settings.}
\textbf{a,} Overall probability of a constructive next user turn after scaffolded versus reference assistant turns; annotations show scaffolded-minus-reference differences.
\textbf{b,} Probability of a constructive next user turn by prior user state and assistant support condition; annotations show scaffolded-minus-reference differences.
\textbf{c,} Probability that the next assistant turn contains scaffolded support by prior user state.
\textbf{d,} \todayrevise{Pooled metadata-adjusted odds ratios} for focal support forms within prior user states. M1 feedback and M4 explaining are shown because Section~\ref{subsec:support_form} identified them as focal positive conversation-level forms; full M1--M6 interaction estimates are reported in Supplementary Table~\ref{tab:prior_state_support_form_interactions}, and operational support-form examples are provided in Supplementary Table~\ref{tab:label_examples}.
Error bars in panels a and b show conversation-cluster bootstrap 95\% confidence intervals over adjacent assistant-to-user pairs; panel d reports adjacent-turn logistic models with conversation-cluster robust standard errors. WC, WildChat; LMSYS, LMSYS Chat; SC, ShareChat.
}
  \label{fig:temporal_coupling}
\end{figure}

\subsection{Support--engagement coupling is state- and form-dependent}\label{subsec:2.5}


The form-level result in the previous section makes the temporal question sharper: if support forms partly mark how the assistant responds to the user's ongoing work, then their adjacent-turn associations should be evaluated after accounting for the user's immediately preceding state.
We therefore examined local adjacent-turn ordering, asking which support forms tended to follow each engagement state in the preceding user turn and whether constructive participation in the subsequent user turn varied across support forms.

At this adjacent-turn scale, scaffolded assistant turns were more likely than reference assistant turns to be followed by constructive engagement in all six task settings (Fig.~\ref{fig:temporal_coupling}a). The next-turn constructive lift was positive in every setting, ranging from +0.27 to +6.21 percentage points. Five of the six adjacent-turn lifts were statistically distinguishable from zero in conversation-cluster bootstrap tests; LMSYS Chat writing was positive but weaker (+0.27 percentage points, 95\% CI -0.01 to +0.56, $p=.061$; Supplementary Table~\ref{tab:key_contrast_significance}). The temporal signal was therefore consistently positive across settings, but heterogeneous in magnitude.

This adjacent-turn coupling was organized by the user's immediately preceding state. State-conditioned contrasts were generally positive but varied across prior constructive, active and passive turns (Fig.~\ref{fig:temporal_coupling}b). The user-to-assistant direction showed the same contingency: constructive and active user turns were more often followed by scaffolded support than passive turns, especially in coding-oriented conversations (Fig.~\ref{fig:temporal_coupling}c). These two directions suggest a coupled sequence rather than a one-way support pattern: engaged user states more often preceded scaffolded support, and scaffolded support was then more often followed by further constructive participation.

We then asked whether the form of support shaped this adjacent-turn coupling once prior user state and adjacent-turn context were modelled jointly. We fitted integrated logistic regressions predicting whether the next user turn was constructive, combining prior user state, assistant-scaffolding descriptors, task and framing variables, turn position and \todayrevise{assistant-model or service-source fixed effects}. The scaffolding block improved fit beyond user state and context (full scaffolding block LR $\chi^2_7=1339.6$, $p<.001$; Supplementary Table~\ref{tab:integrated_scaffolding_block_tests}), and prior-state $\times$ support-form interactions improved fit further (\todayrevise{pooled metadata fixed effects}: LR $\chi^2_{18}=297.3$, $p<.001$; Supplementary Table~\ref{tab:prior_state_support_form_interactions}). \todayrevise{The corresponding pooled coefficient estimates are reported in Supplementary Table~\ref{tab:integrated_regression}.} These tests show that the immediate support--engagement association was not reducible to the user's preceding state alone; it depended jointly on what the user had just made visible and how the assistant supported the next step.

Figure~\ref{fig:temporal_coupling}d highlights feedback-like support and explaining, the two support forms that carried the clearest positive signal in the conversation-level analysis; complete M1--M6 interaction estimates and operational support-form examples are reported in the Appendix (Supplementary Tables~\ref{tab:prior_state_support_form_interactions} and \ref{tab:label_examples}). Explaining provided the more stable adjacent-turn signal, predicting constructive follow-up after prior constructive, active and passive user turns (ORs 1.72, 2.13 and 2.88, respectively; all $p<.001$). Feedback showed a different immediate role: its positive next-turn association was concentrated after prior passive turns, rather than after turns in which users were already active or constructive. This divergence clarifies the conversation-level result: feedback marked exchanges in which user attempts were available for evaluation and repair, whereas explanation more consistently provided material for the user's immediate next move. Support form therefore mattered through its position in the unfolding user--assistant sequence, not as a context-free pedagogical category.

Taken together, the temporal analyses locate support--engagement coupling in adjacent-turn sequences. Constructive participation was most clearly organized where the user's visible state, the assistant's support condition and the form of support were considered together. Learning-oriented engagement in everyday human--LLM interaction is therefore assembled in situated next-turn sequences in which the user's visible work and the assistant's support form jointly shape whether constructive participation continues.

%% file: tables/table3.tex
\begin{table*}[t]
\small
\centering
\caption{
\textbf{Dataset and behavioural overview.}
Summary statistics for the WildChat, LMSYS Chat and ShareChat task settings analysed in the study. Constructive engagement is reported as the deepest observable level within cognitive engagement and serves as the focal user-side measure in subsequent analyses. \revise{Cognitive overall denotes any passive, active or constructive user engagement; constructive level denotes the constructive subset.}
}

\label{tab:dataset_overview}

\setlength{\tabcolsep}{4pt}
\renewcommand{\arraystretch}{1.1}

\resizebox{\textwidth}{!}{%
\begin{tabular}{lrrrcccc}
\toprule
& \multicolumn{3}{c}{Sample size}
& \multicolumn{1}{c}{User framing}
& \multicolumn{1}{c}{Assistant support}
& \multicolumn{2}{c}{Cognitive engagement} \\
\cmidrule(lr){2-4}
\cmidrule(lr){5-5}
\cmidrule(lr){6-6}
\cmidrule(lr){7-8}
Setting
& Conversations
& \makecell{User\\turns}
& \makecell{Assistant\\turns}
& \makecell{Intentional\\conv. (\%)}
& \makecell{Scaffolded\\turns (\%)}
& \makecell{Overall\\(\%)}
& \makecell{Constructive\\level (\%)} \\
\midrule
WildChat coding  & 31,878 & 124,073 & 124,623 & 37.13 & 51.03 & 50.21 & 9.23 \\
WildChat writing & 39,534 & 163,705 & 164,824 & 25.80 & 23.53 & 16.34 & 2.22 \\
LMSYS coding     & 32,114 & 107,065 & 105,174 & 41.49 & 29.07 & 45.26 & 5.44 \\
LMSYS writing    & 21,023 & 77,906  & 76,279  & 20.64 & 19.43 & 15.30 & 1.53 \\
ShareChat coding & 2,481  & 11,541  & 11,522  & 45.63 & 50.67 & 45.00 & 15.17 \\
ShareChat writing & 1,539 & 7,395   & 7,363   & 29.69 & 22.42 & 33.05 & 5.18 \\
\midrule
Total / pooled   & 128,569 & 491,685 & 489,785 & 32.12 & 31.70 & 31.94 & 4.93 \\
\bottomrule
\end{tabular}%
}

\vspace{2mm}
\begin{minipage}{0.98\textwidth}
\footnotesize
\textit{Note.} User framing is reported as the percentage of conversations with explicit learning-oriented framing. \revise{Cognitive and constructive engagement are percentages of user turns;} scaffolded support is computed over assistant turns. \revise{Assistant turns refer to AI/LLM responses, so user-side and assistant-side percentages use different role-specific denominators.}
\end{minipage}
\end{table*}

%% file: sections/03-discussion.tex
\section{Discussion}

\subsection*{Everyday human--LLM interaction as an informal learning ecology}

By tracing engagement in large-scale public logs, this study shows that everyday human--LLM interaction can be analysed as an empirical ecology of informal learning. These conversations are organized around users' own tasks, yet they contain behavioural traces of learning-oriented work. Section~\ref{subsec:2.1} establishes the depth structure of this participation: cognitive engagement appeared in nearly one third of user turns, whereas constructive engagement appeared in about five percent. This distinction matters because broad participation and constructive sense-making mark different depths of involvement in the exchange. Broad participation shows that users acknowledge, apply or extend task material; constructive engagement marks the more selective moments in which users make their reasoning visible by articulating constraints, testing implications, revising assumptions or extending explanations.
In this way, informal learning becomes observable not as a declared intention, but as a pattern of participation within ordinary interaction.

The contextual results in Section~\ref{subsec:2.2} show how this participation was organized. Explicit learning-oriented framing made constructive engagement more common, but task-oriented exchanges also contained constructive turns. Task ecology shaped which forms of sense-making became visible: coding conversations often exposed failure, constraints and tests, whereas writing conversations made revision, adaptation and evaluative choice visible when users brought those concerns into the exchange. Sustained conversations added temporal opportunity, allowing earlier responses to become points of return, comparison and revision.
\latestrevise{
Beyond these factors, we have also tested the impact of LLM model (for example, GPT-4- versus GPT-3.5-class systems) or family (for example, GPT or Gemini) selection with fixed-effect sensitivity checks.
The results reveal that our main findings in Sections~\ref{subsec:2.1}--\ref{subsec:2.5} held across different model conditions (refer to Supplementary Section~\hyperref[supp:a5]{A.5}).}
Together, these findings define everyday human--LLM interaction as an informal learning ecology: user orientation, task affordances and interaction depth shape when ordinary problem solving becomes an occasion to build, test and extend understanding.

\subsection*{Scaffolding as situated support}

The assistant-side findings show how support enters this ecology. As shown in Section~\ref{subsec:2.3}, conversations containing scaffolded support had richer constructive participation across all six task settings, and this association remained positive after adjustment for observed conversation-level context. This pattern is most informative when scaffolded support is read as a relation between the assistant response and the user's next possible action. In everyday task-oriented LLM use, an answer can do more than provide a usable deliverable: it can expose a rationale, error, mechanism, constraint or partial solution that remains available for inspection. This view extends scaffolding theory, which emphasizes contingent assistance and the gradual transfer of responsibility~\cite{wood1976role,van2010scaffolding,reiser2004scaffolding}, into general-purpose AI interaction, where task completion and opportunities for sense-making are interleaved within the same exchange.

The support-form results specify what this situated support looked like in the logs. As shown in Section~\ref{subsec:2.4}, feedback-like support and explanation were most consistently aligned with constructive engagement. Feedback makes a prior attempt evaluable: it turns a user's partial solution, diagnosis or draft into something that can be repaired, refined or extended~\cite{hattie2007power,shute2008focus,yin2025chatbotfeedback}. Explanation makes reasons and mechanisms inspectable, giving users material for comparison, self-explanation and transfer~\cite{chi1989self,renkl1997learning}. Other forms, including instruction, modelling and questioning, supported task progress, examples and clarification, with more task-contingent alignment to visible constructive elaboration. The form-level contrasts point to the warrant that support attaches to generated content: some forms organize action, whereas others make visible the grounds on which an answer, attempt or mechanism can be used. For general-purpose AI assistance, this shifts scaffolding from the surface form of a response to the epistemic status of the output it produces: fluent answers become learning-relevant when they are also judgeable answers, with grounds, limits or failure points that can be inspected and adapted within the task.

\subsection*{Learning continuity across formal and everyday contexts}

The temporal findings clarify what everyday AI assistance can add to a broader learning system. As shown in Section~\ref{subsec:2.5}, support--engagement coupling was most interpretable in adjacent user--assistant--user sequences, where the user's preceding state and the assistant's support form shaped whether constructive participation continued. This scale gives everyday AI use a role that differs from formal instruction. Formal learning environments organize progression through curricula, sequencing, assessment, shared standards and opportunities to abstract beyond the immediate task. Everyday AI assistance organizes re-entry: it brings explanations, feedback and prior reasoning back into the moments where people apply knowledge under practical constraints.

This distinction matters because much learning depends on whether knowledge can be reactivated when it is needed. A concept explained in a course, tutorial or earlier exchange becomes educationally consequential when it can be recognized again in a new problem, adapted to a different constraint or used to diagnose a breakdown. Human--LLM exchanges provide occasions for that re-entry. A user encounters a concrete difficulty; the assistant supplies a rationale, comparison or correction; the user can then relate that material to the task at hand. The value of such episodes lies in making knowledge usable within action, where understanding is tested against the demands of a real problem.

This complementarity suggests a different role for personal AI assistants in education. Schools, courses and curricula provide progression, accountability and shared criteria. AI assistants can provide continuity across contexts: carrying explanations from one task to another, preserving traces of prior attempts, helping users notice recurring uncertainties and supporting consolidation after repeated encounters with a concept. In this role, lifelong AI assistance can function as a connective layer between structured learning and everyday practice, giving memory and personalization to the informal learning that emerges while people work.

\subsection*{AI-mediated work and the social organization of expertise}

The broader implication is that AI-mediated work may reshape the pathways through which expertise is formed. Much professional learning develops through participation in the practices of a domain: proposing first attempts, encountering error, receiving feedback, revising choices, explaining decisions and seeing consequences unfold~\cite{brown1989situated,collins2018cognitive}. These activities are often treated as parts of production, yet they also carry an apprenticeship function. This study provides an empirical entry point into this apprenticeship function by making such participation observable in ordinary AI-mediated work and showing that it emerges selectively under different interactional conditions. As LLMs become embedded in writing, coding, search and problem solving, the formation of expertise therefore becomes tied to how these environments organize participation in the practices through which judgment is built.

AI also changes the repertoire of capacities through which people enter expert practice. When systems can generate fluent drafts, code, explanations and recommendations, expertise may depend increasingly on formulating good problems, recognizing relevant principles, interrogating generated solutions, judging evidence, revising under constraints and remaining accountable for consequences. These capacities extend beyond generic AI literacy; they are forms of situated judgment that develop through repeated participation in domain work. The task- and support-contingent patterns reported above give this claim an interactional basis: opportunities to exercise such judgment are organized within the interaction itself. They also identify a model-design opportunity: sustaining users' cognitive participation can become an explicit objective of model behaviour alongside task success, with the process measures developed here providing a way to evaluate that objective. Schools, workplaces and platforms will therefore shape future expertise through the routines they build around AI: what novices practise, where feedback enters, how responsibility is assigned and how mistakes become occasions for development.

A science of everyday AI learning should therefore study cognitive apprenticeship as a distributed property of sociotechnical systems. The process measures developed in this paper provide one basis for this agenda: they make the allocation of cognitive participation observable at scale and offer a common basis for comparing models, interfaces and institutional routines. The outcomes of interest therefore extend beyond immediate productivity or single-session engagement to the formation of future practitioners: whether repeated AI-mediated routines cultivate judgment, autonomy and domain competence over time. The social stakes are distributional. Intelligent assistance can broaden access to expert-like feedback, explanation and revision, but the developmental value of that access will depend on institutional design: who participates in diagnosis, who practises revision, who observes consequences and who is asked to justify decisions. The long-term educational significance of AI will depend on how societies organize these pathways into expertise, and on whether AI-mediated work sustains broad participation in the practices through which judgment, autonomy and domain competence are formed.


\subsection*{Limitations}


Several boundaries define the interpretation of these findings. Public conversational logs enable large-scale analysis of heterogeneous everyday interactions, but they characterize population-level patterns rather than longitudinal learner histories and do not capture users' prior knowledge, motivation, goals outside the conversation or later learning outcomes. The engagement labels provide a consistent, theory-grounded operationalization of learning-oriented participation in naturalistic conversations, while remaining behavioural indicators rather than direct evidence of retention, transfer or skill development. Assistant support was observed rather than assigned, capturing the naturally occurring covariation and sequencing central to our ecological analysis but limiting causal inference. Finally, the focus on English-language coding and writing limits generalization, although both are prominent domains of AI-mediated knowledge work and provide an informative contrast across three corpora. Evidence from other domains, languages, interfaces and institutional settings would further clarify the generality of these patterns.



\subsection*{Conclusion}

As LLMs become infrastructure for everyday work and study, the central educational question extends beyond how individuals learn from intelligent systems to how learning opportunities are organized in AI-mediated societies~\cite{burton2024collective}. This study shows that ordinary human--LLM interaction already contains measurable behavioural signatures of informal learning, and that deeper constructive engagement emerges selectively through task ecologies, support forms and short interactional sequences. \latestrevise{The broader implication is that AI-mediated work should be assessed not only by the productivity gains it yields or the cognitive effort it displaces, but also by how it expands, constrains or redistributes opportunities to engage in the cognitive work through which people develop knowledge, skills and judgment.} The long-term significance of AI may therefore depend not only on the knowledge or productivity it delivers, but also on how societies organize learning opportunities: who encounters them, how often they arise and whether institutions sustain them over time.





%% file: sections/04-methods_reorganized.tex
\section{Methods}\label{sec6}

\subsection{Study design and analytic scope}

This study analyses naturalistic public human--LLM conversations as
observational digital traces of everyday work and study, following the logic of
large-scale digital-trace and computational social-science analyses~\cite{lazer2009computational,salganik2017bitbybit}.
We use three public conversational corpora: WildChat-4.8M~\cite{allenai2025wildchat48m,zhao2024wildchat}
(public non-toxic release, reported below as WildChat), LMSYS Chat-1M~\cite{zheng2023lmsyschat1m} (reported below as LMSYS Chat) and ShareChat~\cite{yan2025sharechat} (strict-English subset, reported below as ShareChat). Within these corpora, we
focus on coding and writing, two prominent AI-mediated knowledge-work ecologies in which users can delegate task completion to the assistant or remain actively involved in diagnosis, evaluation and revision. Usage and workplace
studies identify writing and software-development tasks as high-salience domains of everyday LLM use and AI-mediated knowledge work~\cite{chatterji2025people,handa2025economic,noy2023experimental,brynjolfsson2025generative}. The two domains also differ in how they make learning-oriented participation
visible: coding often externalizes uncertainty through errors, constraints, execution failures and iterative testing~\cite{barke2023grounded,keuning2018systematic},
whereas writing often involves drafting, revision and evaluative choice whose rationale may remain implicit unless users articulate it~\cite{flower1981cognitive,sommers1980revision}.

Human--LLM conversations are naturally occurring records of user--assistant collaboration.
We analyse them at three scales: conversations, turns and adjacent user--assistant--user sequences. User
turns provide engagement-depth measures, assistant turns provide support measures and adjacent sequences provide local ordering. The analyses estimate where learning-oriented engagement is visible, where constructive engagement is
concentrated and how scaffolded support is associated with further constructive participation across conversation-level, support-form and adjacent-turn scales. Dataset and task ecology are treated as substantive dimensions of the interaction
environment. The claims are therefore about observable behavioural signatures and interactional organization in public logs, with causal and durable-learning questions reserved for experimental or longitudinal designs~\cite{hernan2020causal}.

\subsection{Corpus construction and task settings}

We constructed the analytic corpus from three public human--LLM conversation releases: WildChat-4.8M~\cite{allenai2025wildchat48m,zhao2024wildchat}, LMSYS Chat-1M~\cite{zheng2023lmsyschat1m} and ShareChat~\cite{yan2025sharechat}. For each source, we organized the public records into role-ordered conversation records with user and assistant turns. We then applied corpus-specific language and minimum-turn screens to retain conversations within the English-oriented analysis scope and with at least four message turns.

Coding- and writing-oriented conversations were identified with LLM-assisted semantic task filters, as detailed in Supplementary Section~\hyperref[supp:a1]{A.1} and Supplementary Table~\ref{tab:filter_classifier_validation}. Retained conversations were then processed with the shared annotation pipeline for conversation-level user framing and turn-level labels for user engagement, scaffolded support, support intent and support form; the codebook, workflow and validation details are reported in Supplementary Sections~\hyperref[supp:a2]{A.2}--\hyperref[supp:a4]{A.4}.

This construction process yielded six analytic task settings: WildChat coding, WildChat writing, LMSYS Chat coding, LMSYS Chat writing, ShareChat coding and ShareChat writing. The final corpus contains 128,569 conversations and 981,470 total turns, including 491,685 user turns and 489,785 assistant turns. WildChat and LMSYS Chat contribute the largest public interaction logs, whereas ShareChat contributes a smaller complementary sample.

Supplementary Table~\ref{tab:corpus_filtering_pipeline} reports the corpus-specific raw-to-final filtering ledger, including source records, eligibility screens, task-filtering steps and final analytic samples. Supplementary Table~\ref{tab:filter_classifier_validation} summarizes validation and audit evidence for task, language, user-framing and turn-level labels.

\subsection{Operational constructs}




\subsubsection{Task ecology, user framing and interaction depth}

Each conversation is represented as an ordered sequence of user and assistant turns together with available metadata. The main contextual variables are data source, task ecology, user framing, topic and language. Task ecology and user framing are analytic variables generated by the corpus-construction and annotation pipeline; topic, language and \todayrevise{assistant-model or service-source metadata} are retained where available. User framing is a conversation-level LLM-assisted label that captures whether the
opening user turn positions the interaction as an occasion for learning or developing understanding, rather than only for completing the immediate task.

Interaction depth summarizes how much exchange unfolded after the initial request. We use turn counts and turn ordering as descriptive measures of opportunity for follow-up, repair and elaboration, rather than as manipulated exposures. Post-answer depth is measured as the number of turns after the first assistant response. When analyses use whether a conversation contains at least one constructive turn, interaction depth is interpreted as a visibility condition: longer exchanges provide more occasions for constructive participation to appear. The corresponding rate-based sensitivity checks are described with the statistical analyses.

\subsubsection{User engagement}

The unit of annotation is the conversational turn. User turns are labelled with a depth hierarchy for cognitive engagement as reflected in the exchange, drawing on engagement theory and the ICAP distinction between passive, active and constructive participation~\cite{fredricks2004school,chi2014icap}.
Passive engagement (\texttt{P}) captures reception, acknowledgement or minimal continuation without substantial transformation. Active engagement (\texttt{A}) captures task-relevant use, follow-up or application of supplied information or task material, such as applying a suggestion, asking for a reformatted artifact or requesting the next step. Constructive engagement (\texttt{C}) captures higher-depth cognitive work in which users generate, revise or extend ideas, explanations or constraints, such as testing an implication, revising a hypothesis or assumption, articulating a constraint, transferring a mechanism or extending an explanation. A user turn labelled \texttt{P}, \texttt{A} or \texttt{C} contributes to the overall cognitive-engagement measure.
Other than these three types of engagement, ICAP also includes interactive engagement, which refers to dialogic co-construction across participants and turns.
Since every record in this study is already a human--assistant exchange, assigning an \texttt{I} (interactive engagement) label at the turn level would mostly capture interaction format rather than learning-oriented engagement depth.
As a result, we do not use ICAP's interactive category as a turn-level user label.
\revise{Operational definitions and plain-language examples for these user-turn labels are provided in Supplementary Table~\ref{tab:annotation_codebook_definitions} and Supplementary Table~\ref{tab:label_examples}. Validation evidence for the annotation workflow is reported separately in Section~\ref{subsec:annotation_workflow_validation} and Supplementary Section~\hyperref[supp:a4]{A.4}.}

Constructive engagement is the focal user-side measure in the main analyses because it identifies where the user contributes explicit reasoning,
testing and revision. We represent it as a turn-level \texttt{C} indicator, a conversation-level count, an indicator for at least one constructive user turn and a
constructive-turn ratio. The corresponding statistical estimands and model specifications are provided in Supplementary Section~\hyperref[supp:a5]{A.5}.

\subsubsection{Assistant support}

Assistant turns are labelled by their role in the user's ongoing work. The top-level support distinction separates reference responses from scaffolded support. Reference
responses primarily provide the requested information, deliverable or stand-alone answer. Scaffolded support captures assistant behaviour that structures, regulates, diagnoses or redirects the user's process while preserving a role for the user's own reasoning. This definition follows the scaffolding tradition, especially contingency, transfer of responsibility and potential fading~\cite{wood1976role,van2010scaffolding,quintana2018scaffolding,reiser2004scaffolding}.

Scaffolded support is the focal assistant-side measure in the main analyses. At the turn level, assistant turns are represented as scaffolded support or reference response. At the conversation level, we use an indicator for whether a conversation contains at least one scaffolded assistant turn. Scaffolded turns are further characterized by support-intent labels (\texttt{I1} metacognitive, \texttt{I2} cognitive and \texttt{I3} affective support) and non-mutually exclusive support-form labels (\texttt{M1} feedback, \texttt{M2} hinting, \texttt{M3} instructing, \texttt{M4} explaining, \texttt{M5} modelling and \texttt{M6} questioning).
The intent labels describe the broad purpose of support, whereas the form labels describe how support is delivered in the turn.
Support-form analyses condition on scaffolded turns and compare scaffolded conversations containing each support-form label with scaffolded conversations that do not contain that label. Because support-form labels are multi-label descriptors, these contrasts trace how particular forms of scaffolding pattern with constructive participation within the broader scaffolded-support ecology.

\revise{Operational definitions and plain-language examples are provided in Supplementary Table~\ref{tab:annotation_codebook_definitions} and Supplementary Table~\ref{tab:label_examples}; case excerpts are provided in Supplementary Tables~\ref{tab:complete_coding_case}--\ref{tab:complete_writing_case}.}
\revise{Aggregate support-intent and support-form profiles are reported in Supplementary Figure~\ref{fig:supp_support_supply}.} Validation metrics for scaffolded support and support-form labels are reported in Supplementary Tables~\ref{tab:label_validation} and \ref{tab:production_label_validation}. The corresponding statistical estimands and model specifications are provided in Supplementary Section~\hyperref[supp:a5]{A.5}.

\subsection{Annotation workflow and validation}\label{subsec:annotation_workflow_validation}

\paragraph{Codebook development.}
The source corpora do not contain native labels for learning-oriented engagement, assistant scaffolding, support intent or support form.
We therefore constructed these labels through an iterative workflow that combined theory-grounded codebook development, independent human review, prompt calibration, parser checks, targeted post-processing and corpus-scale LLM-assisted labelling~\cite{gilardi2023chatgpt,ziems2024can}.
The codebook was derived from engagement theory, the ICAP framework and scaffolding theory~\cite{fredricks2004school,chi2014icap,wood1976role,van2010scaffolding,quintana2018scaffolding,reiser2004scaffolding}.
Two authors
independently labelled overlapping validation cases, discussed disagreements and produced human-confirmed labels for validation analyses.

\paragraph{Corpus-scale labelling.}
After the codebook, prompt templates and structured output schemas stabilized, the frozen annotation workflow was applied to the analytic corpora. The workflow used schema and parser checks, documented post-processing rules for recurring boundary errors and retained annotation metadata for provider, deployment family, API version, prompt version, retry policy and parser checks. WildChat labels reflect the targeted update pass applied after validation, including post-labelling WildChat review samples drawn from completed labelled pools; LMSYS Chat and ShareChat were then processed with the same workflow. Operational definitions, examples, de-identified case excerpts and reproducibility details are provided in Supplementary Table~\ref{tab:annotation_codebook_definitions}, Supplementary Tables~\ref{tab:label_examples}--\ref{tab:complete_writing_case} and Supplementary Sections~\hyperref[supp:a3]{A.3}--\hyperref[supp:a4]{A.4}.

\paragraph{Human validation.}
Human validation is reported in three layers. Human--human agreement assesses codebook operability; corpus-scale label audits compare final LLM-assisted labels with human-confirmed labels; and post-update audits check recurring residual boundary errors.
Agreement is summarized with per-label F1, Matthews correlation coefficient and Gwet's AC1 because the labels are sparse, prevalence-imbalanced and sometimes multi-label~\cite{matthews1975comparison,gwet2008computing}.
Overall, the validation checks supported the focal labels used for the main claims, including intentional user framing, constructive engagement, scaffolded support and support-form labels.
Full sampling designs, review-round sizes, post-labelling checks and per-label metrics are reported in Supplementary Section~\hyperref[supp:a4]{A.4} and Supplementary Tables~\ref{tab:label_validation}--\ref{tab:wildchat_post_labelling_review}. These tables distinguish codebook-operability checks from LLM-assisted label agreement.
Prevalence estimates in the Results are computed from the full corpus-scale labels.

\subsection{Statistical analyses}

\paragraph{Descriptive and contextual estimates.}
\revise{Descriptive analyses estimated engagement prevalence, scaffolding prevalence, passive/active/constructive composition, explicit affective-expression rates and interaction-depth summaries within each task setting (Sections~\ref{subsec:2.1}--\ref{subsec:2.2}).}
\revise{Contextual analyses compared observed user framing, task ecology and interaction-depth groups in Section~\ref{subsec:2.2}. We first fitted a conversation-level logistic model predicting whether a conversation contained at least one constructive user turn. Because the length pattern in Fig.~\ref{fig:engagement_ecology}c uses the same any-event outcome, we also fitted a rate-based sensitivity model that accounts for the number of user turns available for observing constructive participation. This sensitivity model was a grouped-binomial model predicting constructive user turns out of total user turns, with user framing, task ecology, length bucket and dataset fixed effects as predictors. Both models are reported in Supplementary Table~\ref{tab:constructive_context_logit}.}

\paragraph{Conversation-level support associations.}
Conversation-level support analyses compared conversations containing at least one scaffolded assistant turn with reference conversations without scaffolded support (Section~\ref{subsec:2.3}).
Constructive-ratio contrasts used turn-weighted ratios within each comparison group, defined as total constructive user turns divided by total user turns, with uncertainty estimated from 1,000 conversation-level bootstrap resamples. Covariate-adjusted models reported in Section~\ref{subsec:2.3} used setting-specific Poisson generalized linear models for raw constructive-turn counts and companion logistic models for whether a conversation contained at least one constructive user turn~\cite{mccullagh1989generalized,agresti2013categorical}.
\revise{The primary Poisson model included conversation length as a covariate; offset-rate and quasi-Poisson sensitivities are reported in Supplementary Section~\hyperref[supp:a5]{A.5} and Supplementary Table~\ref{tab:fig3_offset_rate_sensitivity}.}
The resulting estimates are interpreted as covariate-adjusted associations in naturalistic conversation logs.

\paragraph{Support-form analyses.}
Support-form analyses conditioned on scaffolded conversations (Section~\ref{subsec:2.4}).
For each support-form label, we compared scaffolded conversations containing that label with scaffolded conversations that did not contain that label, and reported turn-weighted constructive user-turn ratio differences. Because support-form labels are multi-label descriptors, these contrasts trace how particular forms of scaffolding pattern with constructive participation within the broader scaffolded-support ecology. Bracket tests in Fig.~\ref{fig:support_form_supply} compare displayed strata for each support form; q values use Benjamini--Hochberg false-discovery-rate adjustment within each displayed family of six support-form comparisons~\cite{benjamini1995controlling}.

\paragraph{Metadata sensitivity checks.}
\todayrevise{Available assistant-model metadata and service-source metadata were used as robustness checks across the Results, rather than as primary explanatory variables. Section-specific implementations and interpretation guidance are reported in Supplementary Section~\hyperref[supp:a5]{A.5} and the corresponding supplementary tables.}

\paragraph{Adjacent-turn analyses.}
Temporal analyses focused on adjacent-turn ordering (Section~\ref{subsec:2.5}). \revise{An adjacent-turn sequence is defined as the immediately preceding user turn, the assistant turn and the immediately following user turn.}
We compared the probability that the next user turn was constructive after scaffolded versus reference assistant turns, estimated the same contrast within prior user states and estimated the probability that the next assistant turn was scaffolded after different prior user states. Adjacent-turn contrasts used 1,000 conversation-cluster bootstrap resamples because multiple turn pairs can come from the same conversation.

Integrated adjacent-turn regressions combined prior user-state features, assistant-scaffolding features, support-form descriptors and adjacent-turn context in one logistic model predicting whether the next user turn was constructive. These models included task and framing variables, turn position and dataset fixed effects.
Standard errors were clustered by conversation. We report broad scaffolded-support models separately from support-form decompositions because support-form labels are descriptors nested within scaffolded support. Nested likelihood-ratio block tests first add broad scaffolded-support presence and then test whether support-form descriptors add signal within scaffolded support. Supplementary Section~\hyperref[supp:a5]{A.5} and Supplementary Table~\ref{tab:setting_model_covariates} give the model specifications, covariate sets and sensitivity checks.

%% file: sections/06-declarations.tex
\section*{Data availability}

This study analyses public human--LLM conversation corpora released by their original data providers: WildChat, LMSYS Chat and ShareChat. Readers should obtain the raw public conversation data from the original releases and comply with the corresponding dataset licences, terms of use and redistribution restrictions. The manuscript does not redistribute verbatim raw conversation logs, user identifiers or linked user histories. The Supplementary includes two de-identified illustrative case excerpts derived from retained conversations and formatted to preserve turn sequence, task structure and annotation decisions while reducing privacy and licence risk.

Derived data sufficient to verify the reported aggregate figures, tables and statistical summaries are archived on Zenodo at \url{https://doi.org/10.5281/zenodo.20995945}. The release includes the numeric values underlying the main figures and Appendix C visual summaries, de-identified conversation-level and adjacent-turn analytic labels without raw message text, table source files, confidence intervals, p values, false-discovery-rate q values, model-output summaries, and provenance documentation.

\section*{Code availability}

Custom code used to export source data and de-identified label tables, compute statistical analyses and generate figures and tables is archived with the derived-data release on Zenodo at \url{https://doi.org/10.5281/zenodo.20995945} and is also available in the GitHub repository at \url{https://github.com/CinderD/Informal-learning-in-everyday-human-LLM-interaction}. \revise{The archived repository includes source-data export scripts, statistical-analysis scripts, figure-generation scripts, table source files and annotation reproducibility summaries covering prompt families, output fields, parser checks, retry logic and rule-based post-processing. API keys, raw message text and non-redistributable corpus files are excluded, so full raw-to-label reruns require user-provided credentials and access to the original public corpora.}

\section*{Ethics statement}

The study is an analysis of public, naturalistic human--LLM conversation datasets.
We used the public source corpora as released by their original providers, with de-identification already applied. Reporting is restricted to aggregate statistics, source-data summaries, validation metrics and two de-identified illustrative case excerpts that preserve turn order and annotation decisions.
Human validation was used to assess annotation quality on reviewed examples and did not involve interaction with the original dataset users.


\section*{Acknowledgements and funding}

This project is partially supported by Microsoft Accelerating Foundation Models Research grant, as well as RGC GRF grant 16218724.


\section*{LLM use statement}

\todayrevise{Large language models were used in four roles. First, LLMs were the systems under study: the analysed data consist of public human--LLM conversations from WildChat, LMSYS Chat and ShareChat. Available assistant-model or service-source identifiers were retained as observational metadata for descriptive and robustness analyses. Second, LLMs were used as measurement tools for task filtering, English-oriented filtering and corpus-scale annotation of user framing, engagement, scaffolding, support intent and support form. These annotation workflows used fixed prompts, schema and parser checks, retry logic, post-processing rules and human validation; the reported reproducibility materials summarize model or deployment families and implementation controls in Supplementary Section~\hyperref[supp:a3]{A.3}. Third, LLMs were used during annotation development to test prompts and inspect boundary cases against human-reviewed examples. Fourth, LLM-based tools assisted with code drafting, debugging, workflow organization and language editing. LLM-generated assistance was not used as a substitute for human validation, audit-label adjudication, statistical decisions or interpretation of the results. No LLM or AI tool is listed as an author, and the authors take responsibility for the final manuscript, analyses and interpretation.}

\section*{Figure source data}
\begin{tabularx}{\textwidth}{@{}>{\raggedright\arraybackslash}p{0.20\textwidth}>{\raggedright\arraybackslash}X@{}}
Main figures & Numeric source-data CSV files are archived in the Zenodo release at \url{https://doi.org/10.5281/zenodo.20995945}. Within the release archive, the main-figure files are \texttt{source\_data/figure2\_source\_data.csv}, \texttt{source\_data/figure3\_source\_data.csv}, \texttt{source\_data/figure4\_source\_data.csv} and \texttt{source\_data/figure5\_source\_data.csv}. \\
Fig.~1 & Conceptual framework figure; no numeric source-data file. \\
Appendix displays and tables & Source files for Appendix C visual summaries, supplementary tables and regression outputs are archived in the same Zenodo release under \texttt{source\_data/}, \texttt{tables/} and \texttt{statistical\_outputs/}. \\
\end{tabularx}


%% file: sections/05_supplementary_revised.tex
\clearpage
\section*{Supplementary information}
\phantomsection\label{supp:info}
Supplementary information is organized to support reproducibility and interpretation. Appendix \hyperref[supp:appendix-a]{A} records the corpus ledger, annotation ontology, corpus-scale annotation workflow, validation evidence and model specifications. Appendix \hyperref[supp:appendix-b]{B} contains the supporting tables and model outputs referenced from the main text and Appendix A. Appendix \hyperref[supp:appendix-c]{C} contains the supporting figures.

\makeatletter
\setlength{\@fptop}{0pt}
\setlength{\@fpsep}{8pt plus 2pt}
\setlength{\@fpbot}{0pt plus 1fil}
\makeatother

\providecommand{\appcomment}[1]{\par\noindent\textcolor{orange!85!black}{\textbf{Comment addressed:} #1}\par}
\providecommand{\apporig}[1]{\par\noindent\textcolor{red}{\textbf{Original:} #1}\par}
\providecommand{\apprev}[1]{\par\noindent\textbf{Revised:} #1\par}

\section*{Appendix A\quad Reproducibility protocol and construct ledger}
\phantomsection\label{supp:appendix-a}
\setcounter{table}{0}
\renewcommand{\thetable}{A\arabic{table}}

\subsection*{A.1\quad Corpus provenance and filtering ledger}
\phantomsection\label{supp:a1}

\revise{Supplementary Table~\ref{tab:corpus_filtering_pipeline} reports the corpus-specific filtering counts from public source records to the final analytic samples.}

\input{tables/table_corpus_filtering_pipeline}

\clearpage
  
\revise{Supplementary Table~\ref{tab:corpus_provenance_artifacts} records the public source links, source counting units and retained metadata fields used during corpus conversion.}

\input{tables/table_corpus_provenance_artifacts}

\revise{Supplementary Table~\ref{tab:filter_classifier_validation} maps each non-native filter or analytic label to its assignment procedure, validation evidence and role in the analyses.}

\input{tables/table_filter_classifier_validation}

\clearpage

\subsection*{A.2\quad Annotation ontology and operational examples}
\phantomsection\label{supp:a2}

\paragraph{Codebook and examples.}
The Methods summarize the role-specific constructs used in the analysis. This section provides the operational codebook, boundary rules and examples needed to reproduce the annotation decisions. Supplementary Table~\ref{tab:annotation_codebook_definitions} gives the formal label definitions; Supplementary Table~\ref{tab:label_examples} gives paraphrased examples without reproducing raw user text; and Supplementary Tables~\ref{tab:complete_coding_case} and \ref{tab:complete_writing_case} provide de-identified illustrative case excerpts. \revise{Support-intent and support-form profiles are visualized in Supplementary Figure~\ref{fig:supp_support_supply}.}

\input{tables/table_annotation_codebook_definitions}

\clearpage

\subsection*{A.3\quad Corpus-scale annotation implementation}
\phantomsection\label{supp:a3}

Implementation details for the annotation workflow are summarized in Supplementary Tables~\ref{tab:prompt_template_summary} and \ref{tab:annotation_implementation_controls}.

\input{tables/table_prompt_template_summary}

\input{tables/table_annotation_implementation_controls}

\clearpage

\subsection*{A.4\quad Human validation and corpus-scale label reliability}
\phantomsection\label{supp:a4}

\paragraph{Validation design.}
Human review served two measurement roles. First, human--human agreement assessed whether trained reviewers could apply the finalized codebook consistently. The coding codebook check combined three validation rounds covering 658 task turns, with 643 turns after deduplication; the writing check used 210 reviewed conversations. Supplementary Table~\ref{tab:label_validation} reports the per-label metrics for this codebook-operability layer. Second, human--LLM audits compared final LLM-assisted corpus-scale labels with human-confirmed labels after item-level review. These audits covered user framing, user engagement and assistant support across the six corpus-by-task settings and are reported in Supplementary Table~\ref{tab:production_label_validation}.

\paragraph{Agreement metrics.}
Agreement is reported with per-label F1, Matthews correlation coefficient (MCC) and Gwet's AC1 because the labels are sparse, prevalence-imbalanced and sometimes multi-label~\cite{matthews1975comparison,gwet2008computing}. \revise{For interpretation, we treated F1 values around 0.70 or higher as acceptable for a sparse boundary label, values around 0.80 or higher as strong, and values around 0.90 or higher as very strong. MCC and Gwet's AC1 were used as complementary prevalence-aware checks, with values around 0.60 indicating substantial agreement and values around 0.80 or higher indicating strong agreement.} In the human--human codebook checks, the focal labels used in the main analyses showed acceptable or better agreement: Constructive user engagement reached F1=0.85 in coding and F1=0.89 in writing, and scaffolded support reached F1=0.87 in coding and F1=0.97 in writing (Supplementary Table~\ref{tab:label_validation}). Complete per-label F1, MCC, Gwet AC1 and support values are reported in Supplementary Table~\ref{tab:label_validation}.

\paragraph{Corpus-scale label audits.}
The consolidated human--LLM audits were stratified across the six corpus-by-task settings. The user-framing audit reviewed 450 first user turns, with 75 cases from each setting. Intentional-framing agreement reached F1=0.857, MCC=0.677 and Gwet AC1=0.671. The user-engagement audit reviewed 600 user turns, with 100 cases per setting; the constructive-versus-non-constructive distinction reached F1=0.922, MCC=0.840 and Gwet AC1=0.837. The assistant-side audit reviewed 180 assistant turns, with 30 cases per setting; scaffolded-support agreement reached F1=0.912, MCC=0.590 and Gwet AC1=0.791, and support-form F1 values ranged from 0.691 to 0.938 across M1--M6. Supplementary Table~\ref{tab:production_label_validation} reports the corresponding precision, recall, accuracy and positive-count fields.

\paragraph{Post-labelling WildChat review sample.}
After the WildChat corpus-scale labelling pass, an additional label-composition check reviewed 100 labelled turns from WildChat coding and 100 labelled turns from WildChat writing, split into 50 user turns and 50 assistant turns per task. Supplementary Table~\ref{tab:wildchat_post_labelling_review} reports the observed corpus-scale label composition of these samples.

\clearpage

\subsection*{A.5\quad Statistical estimands and model specifications}
\phantomsection\label{supp:a5}

\paragraph{Notation.}
This subsection gives the statistical notation and model specifications used for the main contrast and regression analyses. Let $i$ index conversations, $s$ task settings, $j$ adjacent assistant--user turn pairs and $g$ user-framing strata. $C_i$ is the count of Constructive user turns, $Y_i$ indicates whether conversation $i$ contains at least one Constructive user turn, $S_i$ indicates at least one scaffolded assistant turn in the conversation and $X_i$ denotes the setting-specific conversation-level covariates listed in Supplementary Table~\ref{tab:setting_model_covariates}. GLM specifications follow standard Poisson and logistic forms~\cite{mccullagh1989generalized,agresti2013categorical}.

\paragraph{Conversation-level models.}
The adjusted count model in Section~\ref{subsec:2.3} is fitted separately within each setting:
\begin{align*}
C_i \mid S_i, X_i &\sim \operatorname{Poisson}(\mu_i),\\
\log(\mu_i) &= \alpha_s + \beta_s S_i + \gamma_s^\top X_i .
\end{align*}
The reported Poisson count ratio is $\exp(\beta_s)$. The companion binary model is
\begin{align*}
\operatorname{logit}\{\Pr(Y_i=1 \mid S_i, X_i)\}
  = \alpha_s + \beta_s S_i + \gamma_s^\top X_i ,
\end{align*}
with $\exp(\beta_s)$ reported as the odds ratio. The user-framing stratified count model uses
\begin{align*}
\log(\mu_i) = \alpha_{s,g} + \beta_{s,g} S_i + \gamma_{s,g}^{\top} X_i^{(-g)},
\end{align*}
where $X_i^{(-g)}$ omits user framing because the model is fitted within stratum $g$.
These models support the Section~\ref{subsec:2.3} adjusted scaffolded-support results in Fig.~\ref{fig:s2_conversation_association}b,c and Supplementary Table~\ref{tab:support_engagement_summary}; setting-specific covariates are listed in Supplementary Table~\ref{tab:setting_model_covariates}.

\paragraph{Rate and context checks.}
The Section~\ref{subsec:2.2} constructive-rate sensitivity is a grouped-binomial logistic model:
\begin{align*}
C_i &\sim \operatorname{Binomial}(T_i,\pi_i),\\
\operatorname{logit}(\pi_i) &=
\alpha + \delta_1 F_i + \delta_2 Q_i + \delta_3 L_i + \delta_4 D_i ,
\end{align*}
where $T_i$ is the number of user turns, $F_i$ user framing, $Q_i$ task ecology, $L_i$ conversation-length bucket and $D_i$ dataset fixed effects. The Section~\ref{subsec:2.3} offset-rate sensitivity uses the same Poisson mean model as above with
\begin{align*}
\log(\mu_i) = \log(T_i) + \alpha_s + \beta_s S_i + \gamma_s^\top X_i .
\end{align*}
Supplementary Tables~\ref{tab:fig3_offset_rate_sensitivity} and \ref{tab:constructive_context_logit} report the estimates.
The grouped-binomial model is the Section~\ref{subsec:2.2} sensitivity for the conversation-length pattern in Fig.~\ref{fig:engagement_ecology}c, whereas the offset-rate Poisson model is the Section~\ref{subsec:2.3} sensitivity for the scaffolded-support count models in Fig.~\ref{fig:s2_conversation_association}b.
\todayrevise{Supplementary Table~\ref{tab:section21_model_source_breakdown} reports a metadata breakdown for the Section~\ref{subsec:2.1} descriptive prevalence measures. Supplementary Tables~\ref{tab:section22_model_source_sensitivity} and \ref{tab:section23_model_source_sensitivity} report parallel metadata sensitivity checks in which dataset fixed effects are replaced by available assistant-model or service-source indicators. Supplementary Table~\ref{tab:section24_support_form_model_source_sensitivity} reports the corresponding Section~\ref{subsec:2.4} support-form sensitivity, restricted to scaffolded conversations and entering the six non-exclusive support-form labels jointly.}

\paragraph{Adjacent-turn models.}
Let $Z_{ij}$ indicate whether the next user turn after assistant turn $j$ is Constructive, $U_{ij}$ encode the immediately preceding user state, $A_{ij}$ encode scaffolded support at assistant turn $j$, $M_{ij}$ encode support-form labels M1--M6 and $R_{ij}$ denote task, framing, turn-position and fixed-effect controls. The integrated adjacent-turn model is
\begin{align*}
\operatorname{logit}\{\Pr(Z_{ij}=1)\}
  = \alpha + \theta^\top U_{ij} + \beta A_{ij}
  + \phi^\top M_{ij} + \eta^\top (U_{ij}\times M_{ij}) + \rho^\top R_{ij}.
\end{align*}
The broad scaffolded-support model omits $M_{ij}$ and $U_{ij}\times M_{ij}$; \todayrevise{metadata sensitivity checks replace dataset fixed effects in $R_{ij}$ with available assistant-model or service-source indicators}.
These models support the Section~\ref{subsec:2.5} adjacent-turn interpretation in Fig.~\ref{fig:temporal_coupling}d and Supplementary Tables~\ref{tab:integrated_regression}--\ref{tab:prior_state_support_form_interactions}; descriptive adjacent-turn lifts in Fig.~\ref{fig:temporal_coupling}a,b are summarized with bootstrap uncertainty in Supplementary Table~\ref{tab:key_contrast_significance}.

\paragraph{Uncertainty and multiplicity.}
Percentage-point contrasts use 1,000 conversation-level bootstrap resamples. Adjacent-turn contrasts use 1,000 conversation-cluster bootstrap resamples. \revise{Primary adjusted conversation-level GLMs use model-based standard errors, with HC1 or quasi-Poisson sensitivities reported where noted;} adjacent-turn regressions use conversation-cluster robust standard errors. Support-form bracket tests use Benjamini--Hochberg false-discovery-rate adjustment within each six-form comparison family~\cite{benjamini1995controlling}.

\paragraph{Metadata robustness checks.}
\latestrevise{Assistant-model metadata in WildChat and LMSYS Chat, and service-source metadata in ShareChat, provided an additional robustness layer for checking whether the reported patterns were stable across observable model and platform variation. These metadata enter the analyses in three ways. For Section~\ref{subsec:2.1}, Supplementary Table~\ref{tab:section21_model_source_breakdown} reports descriptive ranges of engagement and scaffolding prevalence across metadata levels within each setting. For Sections~\ref{subsec:2.2}--\ref{subsec:2.4}, pooled sensitivity models refit the dataset-fixed-effect specifications after replacing dataset fixed effects with assistant-model or service-source fixed effects (Supplementary Tables~\ref{tab:section22_model_source_sensitivity}--\ref{tab:section24_support_form_model_source_sensitivity}). For Section~\ref{subsec:2.5}, metadata-adjusted adjacent-turn regressions are reported in pooled form, with nested block tests and prior-state by support-form interactions used to evaluate the local sequence pattern (Supplementary Tables~\ref{tab:integrated_regression}--\ref{tab:prior_state_support_form_interactions}).}
\latestrevise{The tables should be read according to their estimand. Descriptive ranges show how prevalence measures vary across available assistant-model or service-source levels. In odds-ratio, count-ratio and rate-ratio tables, values above 1 indicate higher constructive engagement associated with the focal predictor, values below 1 indicate lower constructive engagement and confidence intervals excluding 1 indicate that the coefficient is statistically distinguishable from the null under that specification. The key robustness signal is preservation of direction and comparable magnitude when dataset fixed effects are replaced by metadata fixed effects. Pooled tables summarize corpus-wide stability; adjacent-turn block tests and interaction models show whether the local sequence pattern remains after user state, support condition, support form and metadata variation are considered together.}

\section*{Appendix B\quad Source tables and model outputs}
\phantomsection\label{supp:appendix-b}
\setcounter{table}{0}
\renewcommand{\thetable}{B\arabic{table}}
\setlength{\floatsep}{20pt plus 2pt minus 2pt}
\setlength{\textfloatsep}{20pt plus 2pt minus 2pt}
\makeatletter
\setlength{\@fpsep}{20pt plus 2pt}
\makeatother

\input{tables/table_label_examples}

\input{tables/table_complete_case_examples}

\input{tables/table_label_validation}

\clearpage

\input{tables/table_production_label_validation}

\clearpage

\input{tables/table_wildchat_post_labelling_review}

\input{tables/table_setting_model_covariates}

\clearpage

\input{tables/table_key_contrast_significance}

\clearpage

\input{tables/table4}

\clearpage

\input{tables/table_fig3_offset_rate_sensitivity}

\input{tables/table_section21_model_source_breakdown}

\input{tables/table_emotional_expression_summary}

\clearpage

\input{tables/table_constructive_rate_context_logit}

\clearpage

\input{tables/table_section22_model_source_sensitivity}

\input{tables/table_section23_model_source_sensitivity}

\clearpage

\input{tables/table_section24_support_form_model_source_sensitivity}

\clearpage

\input{tables/table_integrated_regression}

\clearpage

\input{tables/table_integrated_scaffolding_block_tests}

\input{tables/table_prior_state_support_form_interactions}

\clearpage

\section*{Appendix C\quad Supporting figures}
\phantomsection\label{supp:appendix-c}
\setcounter{figure}{0}
\renewcommand{\thefigure}{C\arabic{figure}}

\begin{figure*}[!htbp]
\centering
\includegraphics[width=0.98\textwidth]{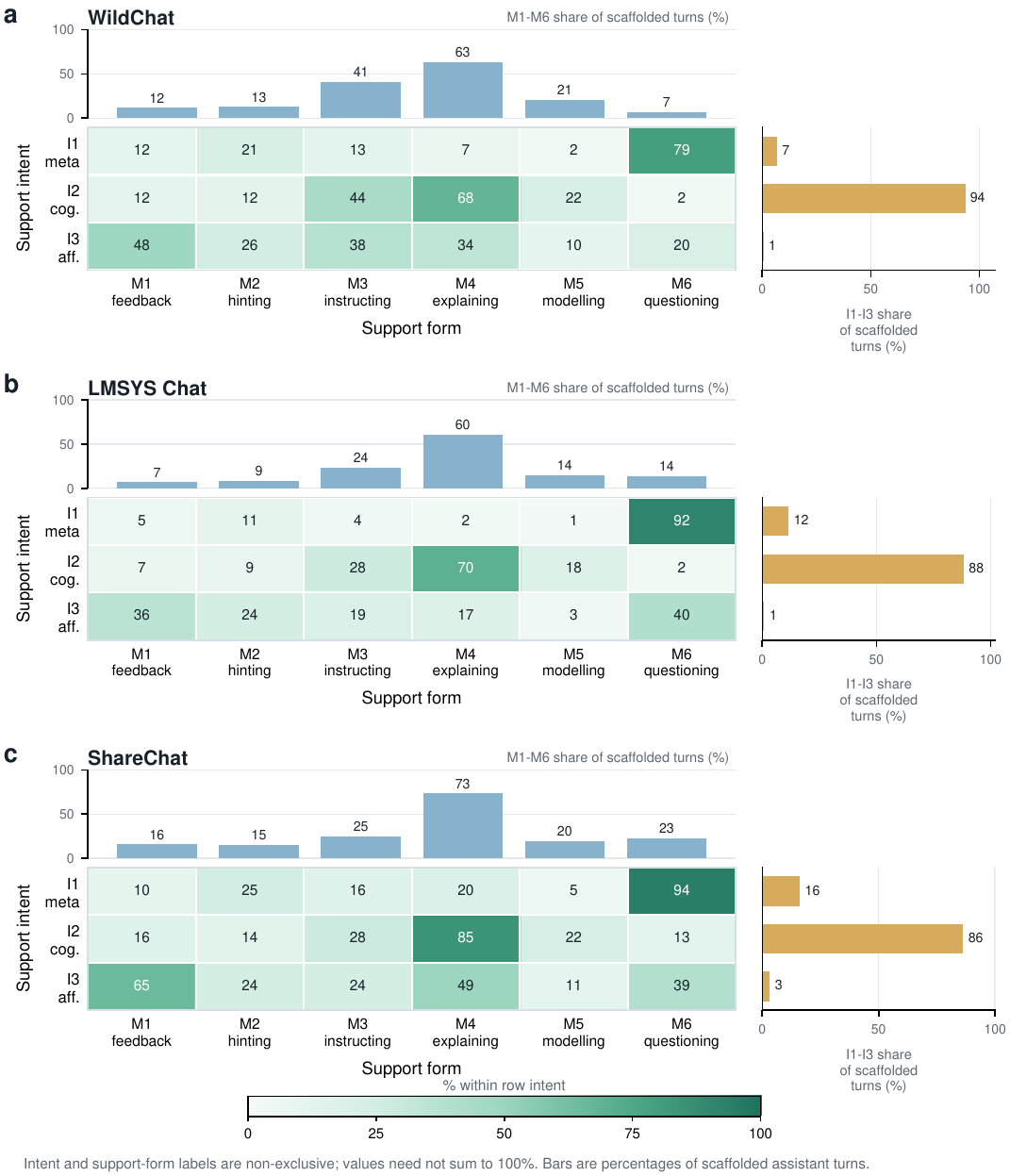}
\caption{\textbf{Support-intent and support-form profiles within scaffolded assistant turns.} \textbf{a--c,} Profiles for WildChat, LMSYS Chat and ShareChat, respectively. In each panel, top bars report the percentage of scaffolded assistant turns containing each support-form label (M1--M6), the central heatmap reports the percentage of turns with each support-intent label (I1--I3) that also contain each support-form label and right bars report the percentage of scaffolded assistant turns containing each support-intent label. \todayrevise{These values describe the composition of scaffolded assistant support; larger values indicate more frequent use of that support intent or form, not a larger estimated effect on user engagement.} Intent and support-form labels are non-exclusive, so values need not sum to 100\%.}
\label{fig:supp_support_supply}
\end{figure*}

\paragraph{Support-form figures.}
\revise{Supplementary Figure~\ref{fig:supp_support_supply} combines corpus-level support-intent and support-form summaries across WildChat, LMSYS Chat and ShareChat.} Operational examples for individual support-form labels are provided in Supplementary Table~\ref{tab:label_examples}.

%% file: tables/table_corpus_filtering_pipeline.tex
\begin{table*}[!htbp]
\fontsize{6.2pt}{7.0pt}\selectfont
\centering
\caption{\textbf{Corpus filtering ledger.} Counts show the corpus-specific steps from public source records to the final coding- and writing-oriented analytic samples. Filter validation evidence is reported separately in Supplementary Table~\ref{tab:filter_classifier_validation}.}
\label{tab:corpus_filtering_pipeline}
\setlength{\tabcolsep}{3pt}
\renewcommand{\arraystretch}{1.08}
\begin{tabularx}{\textwidth}{>{\raggedright\arraybackslash}p{0.11\textwidth}>{\raggedright\arraybackslash}p{0.20\textwidth}>{\raggedright\arraybackslash}p{0.18\textwidth}>{\raggedright\arraybackslash}p{0.31\textwidth}>{\raggedright\arraybackslash}X}
\toprule
Corpus
& Public source records
& Eligibility screen
& Task and language filtering
& Final analytic sample \\
\midrule
WildChat
& WildChat-4.8M public release.
& English-oriented records with at least four message turns.
& Keyword prefilter followed by LLM-assisted semantic filters: \texttt{coding\_english} or \texttt{writing\_english}, score $\geq .70$. One coding record with missing \texttt{learning\_intent} was excluded.
& 31,878 coding; 39,534 writing. \\
\addlinespace
LMSYS Chat
& LMSYS Chat-1M public release: 1,000,000 conversations.
& 777,453 English conversations; 236,372 English conversations with at least four turns.
& LLM-assisted semantic filters applied to the English pool with the minimum-turn screen: 32,114 coding; 21,023 writing.
& 32,114 coding; 21,023 writing. \\
\addlinespace
ShareChat
& ShareChat public release: 142,808 public conversations and 660,293 public turns. Local conversion used 1,199,899 message rows and 129,584 platform-url conversations.
& 69,915 conversations with at least four message turns.
& LLM-assisted semantic filters selected 3,494 coding and 1,663 writing conversations; the final strict-English screen retained 2,481 coding and 1,539 writing.
& 2,481 coding; 1,539 writing. \\
\bottomrule
\end{tabularx}

\vspace{1mm}
\begin{minipage}{0.96\textwidth}
\fontsize{6.2pt}{7.0pt}\selectfont
\textit{Note.} Counts use corpus-specific source units. The minimum-turn screen uses message count, not user--assistant round count. For ShareChat, public turns and local message rows are different units. ``Strict-English'' refers only to the additional ShareChat final-language screen.
\end{minipage}
\end{table*}

%% file: tables/table_corpus_provenance_artifacts.tex
\begin{table}[!htbp]
\centering
\caption{\textbf{Public source corpora and retained metadata.} The table records the source releases, source-level counting units and corpus-specific metadata retained for corpus construction and robustness checks. Filtering counts are reported separately in Supplementary Table~\ref{tab:corpus_filtering_pipeline}.}
\label{tab:corpus_provenance_artifacts}
\fontsize{7.4pt}{8.5pt}\selectfont
\setlength{\tabcolsep}{3pt}
\renewcommand{\arraystretch}{1.15}
\begin{tabularx}{\textwidth}{>{\raggedright\arraybackslash}p{0.14\textwidth}>{\raggedright\arraybackslash}p{0.27\textwidth}>{\raggedright\arraybackslash}p{0.25\textwidth}>{\raggedright\arraybackslash}X}
\toprule
Corpus & Public source link & Source unit and local conversion note & Corpus-specific metadata retained \\
\midrule
WildChat & WildChat-4.8M public release: \url{https://huggingface.co/datasets/allenai/WildChat-4.8M}. & Public human--LLM conversations were converted to ordered user and assistant messages before task filtering. & Available language/topic fields and user-facing model metadata where available. \\
\addlinespace
LMSYS Chat & LMSYS Chat-1M public release: \url{https://huggingface.co/datasets/lmsys/lmsys-chat-1m}. & The public release contains one record per chat conversation; retained records were converted to ordered user and assistant messages after the English and minimum-turn screens. & Language-filter status, available moderation/redaction fields and model names parsed from conversation identifiers. \\
\addlinespace
ShareChat & ShareChat public release associated with Yan et al.~\cite{yan2025sharechat}: \url{https://huggingface.co/datasets/tucnguyen/ShareChat}. & The public dataset card reports conversations and public turns. Local conversion used message rows and platform URL as the conversation key before deriving user and assistant turns. & Platform URL conversation keys, source platform or assistant-family fields and final conversation-level language flag. \\
\bottomrule
\end{tabularx}

\vspace{1mm}
\begin{minipage}{0.94\textwidth}
\footnotesize
\textit{Note.} This table records source access and retained metadata, not analysis results. All analytic tables also retain internal conversation and turn identifiers, message order, task-filter outputs, user-framing labels and turn-level annotation outputs. Raw conversation text remains governed by the original dataset releases.
\end{minipage}
\end{table}

%% file: tables/table_filter_classifier_validation.tex
\begin{sidewaystable}[p]
\fontsize{7.2pt}{8.4pt}\selectfont
\centering
\caption{\textbf{Filter and label validation map.} The table summarizes how each non-native filter or analytic label was assigned, what validation or audit evidence supports it and where it is used in the analyses. Filtering counts are reported in Supplementary Table~\ref{tab:corpus_filtering_pipeline}; full agreement metrics are reported in Supplementary Tables~\ref{tab:label_validation} and \ref{tab:production_label_validation}.}
\label{tab:filter_classifier_validation}
\setlength{\tabcolsep}{4pt}
\renewcommand{\arraystretch}{1.12}
\begin{tabularx}{\textheight}{>{\raggedright\arraybackslash}p{0.15\textheight}>{\raggedright\arraybackslash}p{0.24\textheight}>{\raggedright\arraybackslash}p{0.25\textheight}>{\raggedright\arraybackslash}X}
\toprule
Variable or filter & How it was assigned & Validation or audit evidence & Role in analysis \\
\midrule
Task ecology: coding versus writing & LLM-assisted semantic task filters selected conversations matching coding-oriented or writing-oriented use after minimum-turn screening. The filters were English-oriented and used a score threshold of 0.70 after keyword prefiltering where applicable. & The filtering ledger records retained counts and selection steps for each corpus (Supplementary Table~\ref{tab:corpus_filtering_pipeline}). In the writing-210 validation artifacts, the conversation-topic field showed high human--LLM agreement (F1=0.973, MCC=0.872, Gwet AC1=0.936; support=210), supporting the topic-label boundary used in the task filter. & Defines the coding- and writing-oriented task settings and the task-ecology covariate used in pooled context models. \\
\addlinespace
User framing: intentional versus unintentional & Conversation-level LLM-assisted \texttt{learning\_intent} annotation marked conversations explicitly oriented toward a goal of learning or developing understanding versus those without explicit learning-oriented framing. & A human-verified audit reviewed 450 first user turns across all six corpus-by-task settings (75 per setting; 50 positive and 25 negative cases according to the corpus-scale labels). Overall intentional-framing agreement was F1=0.857, MCC=0.677 and Gwet AC1=0.671; complete metrics are reported in Supplementary Table~\ref{tab:production_label_validation}. & Used as the observed user-framing variable in Fig.~\ref{fig:engagement_ecology} and the contextual models; prevalence estimates come from the full corpus-scale labels. \\
\addlinespace
Language filtering & WildChat and LMSYS Chat used English-oriented source or semantic filters. ShareChat received an additional strict-English screen after task filtering, retaining only records with final conversation-level language flag \texttt{English}. & ShareChat strict-English retained 2,481 of 3,494 coding conversations and 1,539 of 1,663 writing conversations after task filtering (Supplementary Table~\ref{tab:corpus_filtering_pipeline}). In the writing-210 validation artifacts, language showed near-perfect agreement (F1=0.998, MCC=0.705, Gwet AC1=0.995; support=210), with MCC lower because nearly all reviewed records were English. & Defines the English-oriented analytic sample; the additional strict-English screen applies only to ShareChat. \\
\addlinespace
Turn-level labels: user engagement and scaffolded support & Corpus-scale annotation used the final codebook, parser checks and post-processing rules. Constructive engagement and scaffolded support are the primary user-side measure and assistant-side measure. & Domain-stratified human--human codebook validation shows strong agreement for constructive engagement (coding F1=0.85; writing F1=0.89) and scaffolded support (coding F1=0.87; writing F1=0.97), with full per-label codebook metrics in Supplementary Table~\ref{tab:label_validation}. A WildChat post-labelling check reviewed 100 labelled turns per task under the observed corpus-scale label distribution (Supplementary Table~\ref{tab:wildchat_post_labelling_review}). Human-confirmed corpus-scale label audits across all six corpus-by-task settings showed acceptable or better reliability for user engagement and assistant labels: the 600-case user audit had constructive-vs-non-constructive F1=0.922, MCC=0.840 and Gwet AC1=0.837; the 180-case assistant audit had scaffolded-support F1=0.912, MCC=0.590 and Gwet AC1=0.791, with support-form F1 values ranging from 0.691 to 0.938 (Supplementary Table~\ref{tab:production_label_validation}). & Provides the turn-level measures used in prevalence, association, support-form and adjacent-turn analyses. \\
\bottomrule
\end{tabularx}
\end{sidewaystable}

%% file: tables/table_annotation_codebook_definitions.tex
\begin{table*}[t]
\fontsize{6.8pt}{7.5pt}\selectfont
\centering
\caption{\textbf{Annotation codebook definitions.} The table defines the labels used to translate learning-science constructs into observable turn-level signals. \revise{Passive, active and constructive are user-engagement depth labels; emotional expression is a parallel descriptive label.} These labels are behavioural annotations of conversation text; they are not measures of durable learning, learner identity or instructional efficacy.}
\label{tab:annotation_codebook_definitions}
\setlength{\tabcolsep}{3pt}
\renewcommand{\arraystretch}{1.02}
\begin{tabularx}{\textwidth}{
>{\raggedright\arraybackslash}p{0.13\textwidth}
>{\raggedright\arraybackslash}p{0.15\textwidth}
>{\raggedright\arraybackslash}p{0.32\textwidth}
>{\raggedright\arraybackslash}X}
\toprule
Family & Label & Definition & Boundary rule \\
\midrule
User engagement & Passive (\texttt{P}) & Reception, acknowledgement or minimal continuation without substantial transformation. & Includes short confirmations or requests to continue; excludes task-relevant follow-up or reasoning. \\
User engagement & Active (\texttt{A}) & Task-relevant action, follow-up, application or manipulation of supplied information or task material. & The user takes a task-relevant next step, such as applying a suggestion, reformulating an artifact or requesting the next step, but does not test, revise or extend an idea. \\
User engagement & Constructive (\texttt{C}) & Higher-depth cognitive work in which the user generates, revises or extends ideas, explanations or constraints. & Requires explicit user reasoning or transformation; not inferred from long text alone. \\
\revise{Emotional descriptor} & \revise{Emotional expression (\texttt{E})} & \revise{Explicit emotion in the user turn, such as frustration, gratitude, anxiety or excitement.} & \revise{Coded in parallel with user engagement; not part of the passive--active--constructive cognitive-depth hierarchy.} \\
\midrule
Assistant support & Scaffolding & Assistant support that structures, diagnoses, regulates or redirects the user's process while leaving cognitive or task responsibility with the user. & Distinguished from reference responses that simply provide an answer, draft or deliverable. \\
Support intent & \texttt{I1} metacognitive & Support aimed at planning, monitoring, reflection or regulation of the user's work. & Captures the purpose of support, not its surface form. \\
Support intent & \texttt{I2} cognitive & Support aimed at concepts, mechanisms, task knowledge or problem-solving content. & Most scaffolded support in this corpus is cognitive in purpose. \\
Support intent & \texttt{I3} affective & Support aimed at encouragement, reassurance or affective regulation. & Used only when affective support is substantively present, not for polite tone alone. \\
\midrule
Support form & \texttt{M1} feedback & Evaluative information about the user's prior attempt, diagnosis or understanding. & Requires a target in the user's preceding contribution. \\
Support form & \texttt{M2} hinting & A cue, prompt or partial path that helps the user proceed without fully resolving the task. & Leaves the next inferential or task step with the user. \\
Support form & \texttt{M3} instructing & Explicit procedural guidance, ordered steps or strategy for doing the task. & More directive than hinting; may still count as scaffolding when it structures process. \\
Support form & \texttt{M4} explaining & A rationale, mechanism or causal account that clarifies why something works or fails. & Requires explanatory content, not merely a restatement of an answer. \\
Support form & \texttt{M5} modelling & A reusable pattern, worked example or demonstration intended to be adapted. & Distinguished from simply giving the final requested product. \\
Support form & \texttt{M6} questioning & A question that advances diagnosis, reflection, planning or problem solving. & Excludes generic clarification questions that do not support the user's process. \\
\bottomrule
\end{tabularx}
\vspace{1mm}
\begin{minipage}{0.96\textwidth}
\footnotesize
\textit{Note.} Supplementary Table~\ref{tab:prompt_template_summary} summarizes the prompt-template families used to apply these definitions, and Supplementary Table~\ref{tab:annotation_implementation_controls} reports the corresponding model, parser, retry and boundary-rule controls.
\end{minipage}
\end{table*}

%% file: tables/table_prompt_template_summary.tex
\begin{table}[!htbp]
\centering
\scriptsize
\caption{\textbf{Functional prompt-template summary for filtering and annotation.} The table summarizes the scientific content of the prompt templates used for corpus construction and turn-level annotation. The archived release includes this prompt-family summary, output-field requirements and implementation controls for parser checks, retry logic and rule-based post-processing; API keys, endpoints and other credentials are excluded.}
\label{tab:prompt_template_summary}
\setlength{\tabcolsep}{4.5pt}
\renewcommand{\arraystretch}{1.16}
\begin{tabularx}{\textwidth}{>{\raggedright\arraybackslash}p{0.15\textwidth}>{\raggedright\arraybackslash}p{0.18\textwidth}>{\raggedright\arraybackslash}p{0.27\textwidth}>{\raggedright\arraybackslash}X}
\toprule
Prompt family & Input unit & Required output fields & Core decision rule \\
\midrule
Semantic task filter & Conversation excerpt and available metadata after keyword prefiltering and minimum-turn screening. & Task label, English-oriented flag where applicable, confidence or score and short rationale. & Retain conversations whose substantive user goal matched coding-oriented or writing-oriented task ecology, rather than an incidental mention of code or writing. \\
\addlinespace
Conversation-level user framing & Opening user turn. & \texttt{learning\_intent}: intentional or unintentional; rationale. & Mark intentional framing (explicit learning intent) only when the opening user turn states a goal of learning, understanding why or how, or developing the user's own capability, not merely task completion. \\
\addlinespace
User-turn engagement & User turn with preceding local context. & \revise{One engagement label: none, passive, active or constructive; separate emotional-expression flag; rationale and boundary flags.} & \revise{Classify the user's observable contribution as acknowledgement, task-relevant follow-up or substantive testing, revision, diagnosis, critique or extension, and separately mark explicit emotional expression.} \\
\addlinespace
Assistant scaffolded support & Assistant turn with preceding user context. & Scaffolded-support flag, support-intent labels, support-form labels and rationale. & Mark scaffolded support when the response helps the user diagnose, test, revise or choose a next step, rather than only delivering a finished answer. \\
\addlinespace
Language and strict-English checks & Conversation text after task filtering. & Conversation-level language flag and parse status. & Retain ShareChat records only when the final conversation-level language flag was English; mixed or non-English records were excluded. \\
\bottomrule
\end{tabularx}
\vspace{2pt}
\begin{minipage}{0.98\textwidth}
\footnotesize \textit{Note.} The table reports the constructs and output fields needed to interpret the labels; implementation controls are summarized in Supplementary Table~\ref{tab:annotation_implementation_controls}.
\end{minipage}
\end{table}

%% file: tables/table_annotation_implementation_controls.tex
\begin{table}[!htbp]
\centering
\scriptsize
\caption{\textbf{Corpus-scale annotation implementation controls.} The table records the operational controls used when applying the annotation workflow to the analytic corpora, including deployment families, structured-output constraints, parser checks, retry handling and boundary-rule processing.}
\label{tab:annotation_implementation_controls}
\setlength{\tabcolsep}{5pt}
\renewcommand{\arraystretch}{1.12}
\begin{tabularx}{\textwidth}{>{\raggedright\arraybackslash}p{0.24\textwidth}>{\raggedright\arraybackslash}X}
\toprule
Component & Implementation detail \\
\midrule
Model deployments & Turn-level user and assistant labels used Azure OpenAI chat-completions deployments from the GPT-5.1 family; final writing-oriented calls used GPT-5.1-chat; LLM-assisted semantic filters used GPT-4o-family deployments. \\
\addlinespace
Structured outputs & JSON outputs were used where supported. Final writing calls used JSON-object or JSON-schema constraints, 60-second request timeouts and maximum completion budgets of 300 tokens for conversation-level labels and 350 tokens for turn-level labels. Final coding and writing annotation calls did not pass an explicit temperature parameter; English checks used temperature 0.0 and semantic topic checks used temperature 0.1. \\
\addlinespace
Parser checks & Parser checks enforced required fields, normalized label variants, flagged malformed outputs and recovered missing fields only when the model explanation and structured output made the mapping unambiguous. \\
\addlinespace
Retry and failure handling & Coding runs used up to five attempts with exponential backoff for rate limits and fixed-delay retry for other transient failures. Writing runs used up to six attempts, disabled SDK-level retries, used exponential backoff with jitter and fell back from JSON-schema response format to JSON-object format when needed. Content-filter blocks returned empty analyses and were logged rather than imputed. \\
\addlinespace
Boundary-rule processing & User-turn rules demoted brief acknowledgements, generic acceptance or administrative follow-ups when no new constraint, test, revision, diagnosis or substantive extension was present, and removed Emotional labels for formulaic politeness. Assistant-turn rules cleared support-form labels when a turn was not scaffolded, removed scaffolded-support labels from complete answers labelled as scaffolding only because they were long or task-relevant and recovered missing support-form fields only when one mapping was unambiguous. \\
\bottomrule
\end{tabularx}
\end{table}

%% file: tables/table_label_examples.tex
\begin{table*}[!htbp]
\footnotesize
\centering
\caption{\textbf{Operational examples for turn-level labels.} Examples are de-identified and paraphrased from validation review patterns; they illustrate the decision rule rather than reproducing raw conversation text.}
\label{tab:label_examples}
\setlength{\tabcolsep}{4pt}
\renewcommand{\arraystretch}{1.16}
\begin{tabularx}{\textwidth}{>{\raggedright\arraybackslash}p{0.16\textwidth}>{\raggedright\arraybackslash}p{0.36\textwidth}X}
\toprule
Construct & Paraphrased turn & Annotation rationale \\
\midrule
\revise{\texttt{None} no visible cognitive engagement} & \revise{``New topic: write a short birthday message for my colleague.''} & \revise{The user starts an unrelated request rather than taking up, applying or reflecting on the current task material.} \\
\addlinespace[2pt]
\texttt{P} passive user engagement & ``Thanks, that makes sense.'' & Minimal uptake or acknowledgement without new reasoning, evaluation or task transformation. \\
\addlinespace[2pt]
\texttt{A} active user engagement & ``Can you convert that answer into a command I can run in my setup?'' & Task-relevant follow-up or manipulation of supplied information or task material, but without elaborating or testing an idea. \\
\addlinespace[2pt]
\texttt{C} constructive user engagement & ``If the cache expires before the asynchronous callback returns, would the same race condition still happen?'' & The user tests an implication, boundary condition or transfer of the prior explanation. \\
\addlinespace[2pt]
Scaffolding & ``First isolate the failing step; that will show whether the issue is parsing or state management.'' & The assistant structures the user's process and leaves diagnostic or revision work with the user. \\
\addlinespace[2pt]
\texttt{M1} feedback & ``Your diagnosis is close; the bug is in the index offset, not the loop boundary.'' & Evaluative feedback on the user's prior attempt or understanding. \\
\addlinespace[2pt]
\texttt{M2} hinting & ``Before changing the model, look at whether the malformed token first appears before or after parsing.'' & A targeted cue points the user toward the next diagnostic distinction, but does not give the full procedure or fix. \\
\addlinespace[2pt]
\texttt{M3} instructing & ``Run the failing case, print the parser output, add a guard for empty fields, then rerun the test.'' & Explicit ordered steps tell the user what actions to take and in what sequence. \\
\addlinespace[2pt]
\texttt{M4} explaining & ``The exception occurs because the event loop is already running, so a nested call cannot acquire it.'' & A rationale or mechanism that clarifies why the issue occurs. \\
\addlinespace[2pt]
\texttt{M5} modelling & ``For this report section, use the pattern: claim, evidence, implication; for example, this delay matters because X, which means Y for the team.'' & A concrete template or worked example demonstrates a reusable form the user can adapt. \\
\addlinespace[2pt]
\texttt{M6} questioning & ``Which constraint matters more here: preserving tone or shortening the paragraph?'' & A question that advances problem solving or reflection. \\
\bottomrule
\end{tabularx}
\end{table*}

%% file: tables/table_complete_case_examples.tex
\definecolor{caseborder}{HTML}{D6DEE6}
\definecolor{casefill}{HTML}{F8FAFB}
\definecolor{caseheadfill}{HTML}{EEF3F6}
\definecolor{caseblue}{HTML}{2F5F83}
\definecolor{caseteal}{HTML}{2F7E72}
\definecolor{caseconstruct}{HTML}{9B4F3A}
\definecolor{caseactive}{HTML}{8A6A1F}
\definecolor{caseref}{HTML}{65707A}
\definecolor{casepale}{HTML}{E8EEF2}
\definecolor{casechipfill}{HTML}{DFE8ED}
\definecolor{caseheaderchip}{HTML}{31576A}
\definecolor{casenote}{HTML}{4D5C66}

\newcommand{\casebadge}[3]{%
  \begingroup
  \setlength{\fboxsep}{1.8pt}%
  \colorbox{#1}{\textcolor{#2}{\scriptsize\strut\textbf{#3}}}%
  \endgroup}
\newcommand{\casecode}[1]{\casebadge{casepale}{caseblue}{#1}}
\newcommand{\userturn}[1]{\casebadge{caseblue}{white}{\(\triangleright\) #1}\par{\scriptsize\textcolor{caseblue}{User}}}
\newcommand{\assistantturn}[1]{\casebadge{caseteal}{white}{\(\diamond\) #1}\par{\scriptsize\textcolor{caseteal}{Assistant}}}
\newcommand{\casecontinued}[1]{{\scriptsize\textcolor{caseref}{#1}}}
\newcommand{\activebadge}{\casebadge{caseactive}{white}{\(\rightarrow\) A active}}
\newcommand{\constructivebadge}{\casebadge{caseconstruct}{white}{\(\uparrow\) C constructive}}
\newcommand{\scaffoldbadge}{\casebadge{caseteal}{white}{\(\ast\) Scaffolded}}
\newcommand{\referencebadge}{\casebadge{caseref}{white}{\(\cdot\) Reference}}
\newcommand{\casekey}[1]{\textcolor{caseconstruct}{\textbf{#1}}}
\newcommand{\supportkey}[1]{\textcolor{caseteal}{\textbf{#1}}}
\newcommand{\casechip}[1]{\casebadge{casechipfill}{caseblue}{#1}}
\newcommand{\casefield}[2]{%
  \casebadge{caseheaderchip}{white}{#1}\hspace{5pt}{\scriptsize\textcolor{caseblue}{\textbf{#2}}}}
\newcommand{\caseheader}[6]{%
  \begingroup
  \setlength{\fboxsep}{5.0pt}%
  \colorbox{caseheadfill}{%
  \begin{minipage}{0.965\linewidth}
  {\small\textcolor{caseblue}{\textbf{#1}}\quad \textbf{#2}}\par
  \vspace{3pt}
  {\scriptsize\textcolor{casenote}{#6}}\par
  \vspace{5pt}
  {\casefield{Setting}{#3}\hspace{12pt}\casefield{User framing}{#4}}\par
  \vspace{3pt}
  {\casefield{Excerpt length}{#5}}
  \end{minipage}}%
  \endgroup}
\newcommand{\caselegend}{%
  \begingroup
  {\scriptsize\textcolor{caseblue}{\textbf{Legend}}\quad
  \constructivebadge\hspace{3pt}\activebadge\hspace{3pt}\scaffoldbadge\hspace{3pt}\referencebadge}\par\vspace{1pt}
  {\scriptsize\textcolor{caseblue}{\textbf{Assistant codes}}\quad
  \casecode{I1/I2} support intent; \casecode{M1--M6} support forms (non-exclusive).}
  \endgroup}

\clearpage
\begingroup
\footnotesize
\setlength{\LTleft}{0pt}
\setlength{\LTright}{0pt}
\setlength{\tabcolsep}{3pt}
\renewcommand{\arraystretch}{1.18}
\begin{longtable}{>{\raggedright\arraybackslash}p{0.11\textwidth}>{\raggedright\arraybackslash}p{0.58\textwidth}>{\raggedright\arraybackslash}p{0.25\textwidth}}
\caption{\textbf{De-identified coding case excerpt with turn-level labels.} A retained WildChat coding conversation excerpt formatted for readability. Badges encode role, user engagement and assistant support labels.}
\label{tab:complete_coding_case}\\
\multicolumn{3}{p{0.96\textwidth}}{\caseheader{Coding case}{wxWidgets menu debugging}
{WildChat coding}
{Unintentional}
{4-turn excerpt}
{Feedback and instructions diagnose a likely cause; a follow-up troubleshooting response uses checks and a request for context to continue debugging.}}\\[5pt]
\multicolumn{3}{p{0.96\textwidth}}{\caselegend}\\[5pt]
\toprule
\textbf{Turn} & \textbf{De-identified excerpt} & \textbf{Label and annotation cue} \\
\midrule
\endfirsthead
\caption[]{\textbf{De-identified coding case excerpt with turn-level labels} (continued).}\\
\toprule
\textbf{Turn} & \textbf{De-identified excerpt} & \textbf{Label and annotation cue} \\
\midrule
\endhead
\bottomrule
\endfoot
\userturn{U1} & ``I have this C++ wxWidgets code to popup a menu but can't make it to work, I got an error that can't popup a \texttt{NULL} menu. How can I do this correctly? ... \texttt{hamburgerMenu = new wxMenu();} ... \texttt{PopupMenu(hamburgerMenu, pos);}'' & \activebadge\par The user asks for help debugging a specific popup-menu error. \\
\addlinespace[4pt]
\assistantturn{A1} & ``Based on the code you provided, it seems that the variable \texttt{bpButtonHamburger} is not assigned to an actual \texttt{wxBitmapButton} object. Therefore, when the \texttt{OnHamburgerClick} function is called, \texttt{bpButtonHamburger} is \texttt{NULL}, which leads to the error you're experiencing.'' & \scaffoldbadge\quad \casecode{I2}\ \casecode{M1}\ \casecode{M3}\ \casecode{M4}\par The reply gives \supportkey{feedback} on the likely fault, \supportkey{instructs} what to change and explains why the handler matters. \\
\addlinespace[3pt]
\casecontinued{A1 cont.} & ``To fix this issue, make sure you initialize \texttt{bpButtonHamburger} with a valid \texttt{wxBitmapButton} instance ... note that the event type in \texttt{OnHamburgerClick} has changed from \texttt{wxContextMenuEvent} to \texttt{wxCommandEvent}, as it is more appropriate for handling button click events.'' & \casecontinued{Same scaffolded response; instruction and explanation continued.} \\
\addlinespace[4pt]
\userturn{U2} & ``No I forgot to add that part but the button was already correctly initialized'' & \activebadge\par The user provides task-relevant clarification that redirects the debugging path. \\
\addlinespace[4pt]
\assistantturn{A2} & ``If the \texttt{bpButtonHamburger} button is already correctly initialized and you're still getting the error that the menu is \texttt{NULL}, there may be some other issue causing the problem. Here are some things you can check: make sure the \texttt{hamburgerMenu} object is not being destroyed; verify that \texttt{OnHamburgerClick} is being called; double-check the button ID and event handler; check for conflicting handlers.'' & \scaffoldbadge\quad \casecode{I1}\ \casecode{M2}\ \casecode{M6}\par The reply gives diagnostic cues and uses \supportkey{questioning/request for context} to continue troubleshooting. \\
\addlinespace[3pt]
\casecontinued{A2 cont.} & ``If none of these suggestions help, please provide more context or the complete code so we can further investigate the issue and provide a solution.'' & \casecontinued{Same scaffolded response; questioning/request for context continued.} \\
\end{longtable}
\endgroup

\clearpage
\begingroup
\footnotesize
\setlength{\LTleft}{0pt}
\setlength{\LTright}{0pt}
\setlength{\tabcolsep}{3pt}
\renewcommand{\arraystretch}{1.18}
\begin{longtable}{>{\raggedright\arraybackslash}p{0.11\textwidth}>{\raggedright\arraybackslash}p{0.58\textwidth}>{\raggedright\arraybackslash}p{0.25\textwidth}}
\caption{\textbf{De-identified writing case excerpt with turn-level labels.} A retained WildChat writing conversation excerpt formatted for readability. Badges encode role, user engagement and assistant support labels.}
\label{tab:complete_writing_case}\\
\multicolumn{3}{p{0.96\textwidth}}{\caseheader{Writing case}{Thesis revision rationale}
{WildChat writing}
{Intentional}
{4-turn excerpt}
{The exchange asks why a report sentence is weak, applies the explanation to a revised claim and receives feedback on the revised reasoning.}}\\[5pt]
\multicolumn{3}{p{0.96\textwidth}}{\caselegend}\\[5pt]
\toprule
\textbf{Turn} & \textbf{De-identified excerpt} & \textbf{Label and annotation cue} \\
\midrule
\endfirsthead
\caption[]{\textbf{De-identified writing case excerpt with turn-level labels} (continued).}\\
\toprule
\textbf{Turn} & \textbf{De-identified excerpt} & \textbf{Label and annotation cue} \\
\midrule
\endhead
\bottomrule
\endfoot
\userturn{U1} & ``I'm revising a short report introduction and want to understand why this sentence sounds vague: `This project is good for users because it improves the system.' What makes it weak, and how should I think about making the claim more specific?'' & \constructivebadge\par The user identifies a writing problem, asks for the rationale behind the weakness and requests a revision principle rather than only a corrected sentence. \\
\addlinespace[4pt]
\assistantturn{A1} & ``The sentence is grammatical, but it is vague because `good for users' and `improves the system' do not say who benefits, what changes or why the change matters. A stronger version should name the user group, the concrete action and the outcome. For example, ask yourself: what task becomes easier, faster or less error-prone?'' & \scaffoldbadge\quad \casecode{I2}\ \casecode{M1}\ \casecode{M2}\ \casecode{M4}\par The reply gives feedback on the draft, explains the weakness and offers a guiding diagnostic question. \\
\addlinespace[4pt]
\userturn{U2} & ``So the problem is not grammar but that the sentence hides the causal link. Would this be stronger: `The project reduces repeated data entry for clinic staff by combining two forms into one workflow'?'' & \constructivebadge\par The user restates the principle, diagnoses the original weakness and tests a revised claim against the explanation. \\
\addlinespace[4pt]
\assistantturn{A2} & ``Yes, that is stronger. It names the audience, the mechanism and the benefit, so the reader can see why the project matters. If this is for a formal report, you could make it more concise: `By combining two forms into one workflow, the project reduces repeated data entry for clinic staff.' '' & \scaffoldbadge\quad \casecode{I1}\ \casecode{I2}\ \casecode{M1}\ \casecode{M4}\par The reply gives feedback on the user's revision and explains why the revised sentence better supports the report claim. \\
\addlinespace[3pt]
\casecontinued{A2 cont.} & ``You can then follow it with evidence, such as time saved or fewer duplicate entries, if the report requires support for the claim.'' & \casecontinued{Same assistant response; feedback and explanation continued.} \\
\end{longtable}
\endgroup

%% file: tables/table_label_validation.tex
\begin{table*}[t]
\scriptsize
\centering
\caption{\textbf{Domain-stratified human--human codebook agreement for labels used in the main analyses.} Coding uses 643 deduplicated task turns from the final three-round validation rerun; writing uses the final writing-210 validation labels. F1, Matthews correlation coefficient (MCC) and Gwet's AC1 are reported per binary label. Rare degenerate labels with no positive cases in a validation set are omitted. Corpus-scale human--LLM label checks are reported in Supplementary Table~\ref{tab:production_label_validation}.}
\label{tab:label_validation}
\setlength{\tabcolsep}{4pt}
\renewcommand{\arraystretch}{1.08}
\resizebox{\textwidth}{!}{%
\begin{tabular}{lllrrrr}
\toprule
Domain & Role & Label & F1 & MCC & Gwet AC1 & Support \\
\midrule
Coding & User & Active (\texttt{A}) & 0.75 & 0.72 & 0.87 & 315 \\
Coding & User & Constructive (\texttt{C}) & 0.85 & 0.82 & 0.90 & 315 \\
Coding & User & Passive (\texttt{P}) & 0.88 & 0.87 & 0.96 & 315 \\
Coding & User & \revise{Emotional expression (\texttt{E})} & 0.83 & 0.82 & 0.97 & 315 \\
Coding & Assistant & Scaffolding & 0.87 & 0.69 & 0.70 & 328 \\
Coding & Assistant & Feedback (\texttt{M1}) & 0.90 & 0.89 & 0.95 & 328 \\
Coding & Assistant & Hinting (\texttt{M2}) & 0.79 & 0.75 & 0.89 & 328 \\
Coding & Assistant & Instructing (\texttt{M3}) & 0.81 & 0.77 & 0.92 & 328 \\
Coding & Assistant & Explaining (\texttt{M4}) & 0.86 & 0.79 & 0.84 & 328 \\
Coding & Assistant & Modelling (\texttt{M5}) & 0.84 & 0.79 & 0.87 & 328 \\
Coding & Assistant & Questioning (\texttt{M6}) & 0.82 & 0.81 & 0.95 & 328 \\
\midrule
Writing & User & Active (\texttt{A}) & 0.73 & 0.68 & 0.81 & 90 \\
Writing & User & Constructive (\texttt{C}) & 0.89 & 0.83 & 0.86 & 90 \\
Writing & User & Passive (\texttt{P}) & 0.80 & 0.81 & 0.99 & 90 \\
Writing & User & \revise{Emotional expression (\texttt{E})} & 0.80 & 0.79 & 0.98 & 90 \\
Writing & Assistant & Scaffolding & 0.97 & 0.80 & 0.93 & 120 \\
Writing & Assistant & Feedback (\texttt{M1}) & 0.89 & 0.80 & 0.80 & 120 \\
Writing & Assistant & Hinting (\texttt{M2}) & 0.82 & 0.70 & 0.71 & 120 \\
Writing & Assistant & Instructing (\texttt{M3}) & 0.71 & 0.67 & 0.86 & 120 \\
Writing & Assistant & Explaining (\texttt{M4}) & 0.84 & 0.70 & 0.70 & 120 \\
Writing & Assistant & Modelling (\texttt{M5}) & 0.94 & 0.91 & 0.95 & 120 \\
Writing & Assistant & Questioning (\texttt{M6}) & 0.96 & 0.96 & 0.99 & 120 \\
\bottomrule
\end{tabular}%
}
\end{table*}

%% file: tables/table_production_label_validation.tex
\begin{table*}[t]
\scriptsize
\centering
\caption{\textbf{Consolidated human--LLM corpus-scale label agreement.} The audits compare final LLM-assisted corpus-scale labels with human-confirmed labels using one-vs-rest binary coding for each label. User-framing, user-engagement and assistant-support audits were stratified across the six corpus-by-task settings, with 450 first user turns, 600 user turns and 180 assistant turns reviewed, respectively. Final human labels were confirmed after item-level review. The table reports audit size, LLM-positive counts, human-positive counts, precision, recall, F1, MCC, Gwet AC1 and accuracy.}
\label{tab:production_label_validation}
\setlength{\tabcolsep}{4pt}
\renewcommand{\arraystretch}{1.08}
\resizebox{\textwidth}{!}{%
\begin{tabular}{lllrrrrrrrr}
\toprule
Role & Label & Audit N & LLM label + & Human + & Precision & Recall & F1 & MCC & Gwet AC1 & Accuracy \\
\midrule
\multicolumn{11}{l}{\textit{Conversation-level framing}} \\
Conversation & Intentional framing & 450 & 300 & 239 & 0.770 & 0.967 & 0.857 & 0.677 & 0.671 & 0.829 \\
\midrule
\multicolumn{11}{l}{\textit{User engagement}} \\
User & Constructive (\texttt{C}) & 600 & 300 & 327 & 0.963 & 0.884 & 0.922 & 0.840 & 0.837 & 0.918 \\
User & Active (\texttt{A}) & 600 & 120 & 111 & 0.850 & 0.919 & 0.883 & 0.856 & 0.935 & 0.955 \\
User & Passive (\texttt{P}) & 600 & 90 & 80 & 0.889 & 1.000 & 0.941 & 0.934 & 0.978 & 0.983 \\
User & None & 600 & 90 & 82 & 0.889 & 0.976 & 0.930 & 0.920 & 0.974 & 0.980 \\
\midrule
\multicolumn{11}{l}{\textit{Assistant scaffolding and support forms}} \\
Assistant & Scaffolding & 180 & 144 & 139 & 0.896 & 0.928 & 0.912 & 0.590 & 0.791 & 0.861 \\
Assistant & Feedback (\texttt{M1}) & 180 & 45 & 51 & 1.000 & 0.882 & 0.938 & 0.918 & 0.945 & 0.967 \\
Assistant & Hinting (\texttt{M2}) & 180 & 43 & 45 & 0.930 & 0.889 & 0.909 & 0.880 & 0.930 & 0.956 \\
Assistant & Instructing (\texttt{M3}) & 180 & 51 & 55 & 0.745 & 0.691 & 0.717 & 0.600 & 0.715 & 0.833 \\
Assistant & Explaining (\texttt{M4}) & 180 & 87 & 110 & 1.000 & 0.791 & 0.883 & 0.772 & 0.747 & 0.872 \\
Assistant & Modelling (\texttt{M5}) & 180 & 48 & 79 & 0.958 & 0.582 & 0.724 & 0.631 & 0.642 & 0.806 \\
Assistant & Questioning (\texttt{M6}) & 180 & 35 & 20 & 0.543 & 0.950 & 0.691 & 0.675 & 0.873 & 0.906 \\
\bottomrule
\end{tabular}%
}
\end{table*}

%% file: tables/table_wildchat_post_labelling_review.tex
\begin{table*}[t]
\small
\centering
\caption{\textbf{WildChat post-labelling review sample composition.} The table reports the label composition of post-labelling review samples drawn from the completed labelled WildChat coding and writing pools under the observed corpus-scale label distribution.}
\label{tab:wildchat_post_labelling_review}
\setlength{\tabcolsep}{5pt}
\renewcommand{\arraystretch}{1.08}
\textit{Label composition in post-labelling review samples.}

\vspace{2pt}
\resizebox{\textwidth}{!}{%
\begin{tabular}{lllrrrrrr}
\toprule
Task & Role & Total turns & Active & Passive & Constructive & No engagement & Reference & Scaffolded \\
\midrule
WildChat coding & User & 50 & 24 & 16 & 6 & 4 & -- & -- \\
WildChat writing & User & 50 & 25 & 2 & 2 & 21 & -- & -- \\
WildChat coding & Assistant & 50 & -- & -- & -- & -- & 32 & 17 \\
WildChat writing & Assistant & 50 & -- & -- & -- & -- & 46 & 4 \\
\bottomrule
\end{tabular}
}
\vspace{2pt}
\parbox{0.92\textwidth}{\footnotesize \textit{Note.} One WildChat coding assistant turn had no support label in the post-labelling review file, so the reference and scaffolded counts sum to 49 for that row.}

\end{table*}

%% file: tables/table_setting_model_covariates.tex
\begin{table*}[p]
\scriptsize
\centering
\caption{\textbf{Covariates used in the Section~\ref{subsec:2.3} adjusted support-association models.} Models were fitted separately within each of the six task settings. In all rows, the focal predictor is whether the conversation contains at least one scaffolded assistant turn.}
\label{tab:setting_model_covariates}
\setlength{\tabcolsep}{5pt}
\renewcommand{\arraystretch}{1.10}
\begin{tabularx}{\textwidth}{>{\raggedright\arraybackslash}p{0.20\textwidth}>{\raggedright\arraybackslash}p{0.21\textwidth}>{\raggedright\arraybackslash}p{0.31\textwidth}>{\raggedright\arraybackslash}X}
\toprule
Model & Outcome & Adjustment variables & Reported estimand or check \\
\midrule
Primary Poisson count model
& Constructive user-turn count per conversation.
& User framing; total turns; emotional-expression ratio; observable error marker; persistence-after-failure marker; high-persistence marker.
& Poisson count ratio in Fig.~\ref{fig:s2_conversation_association}b; conversation length enters as total turns, not as an offset. \\
\addlinespace
Companion logistic model
& Whether the conversation contains at least one constructive user turn.
& User framing; total turns.
& Logistic odds ratio in Fig.~\ref{fig:s2_conversation_association}b. \\
\addlinespace
User-framing-stratified Poisson model
& Constructive user-turn count within intentional or unintentional conversations.
& Total turns; emotional-expression ratio; observable error marker; persistence-after-failure marker; high-persistence marker.
& Same count model after stratifying by framing; user framing is omitted because it is fixed within stratum. \\
\addlinespace
Offset-rate sensitivity
& Constructive user-turn count per conversation with \texttt{log(user turns)} as an exposure offset.
& Same observed context as the primary Poisson model, with user turns represented as an offset rather than as an ordinary length covariate.
& Checks whether the scaffolded-support association persists when modelling a constructive-turn rate. \\
\bottomrule
\end{tabularx}
\vspace{2pt}
\parbox{0.94\textwidth}{\footnotesize \textit{Note.} Task ecology is fixed within each setting because the models are fitted separately for WC coding, LMSYS coding, SC coding, WC writing, LMSYS writing and SC writing. Task ecology is used only in pooled context models such as Supplementary Table~\ref{tab:constructive_context_logit}. Error and persistence markers are most interpretable in coding-oriented contexts but are retained as observed conversation-level context when present.}
\end{table*}

%% file: tables/table_key_contrast_significance.tex
\begin{table*}[p]
\scriptsize
\centering
\caption{\textbf{Uncertainty estimates for key support--engagement contrasts.} Scaffolded conversations had higher constructive ratios and greater post-answer depth in all six settings; adjacent-turn constructive lifts were positive in all six settings, with LMSYS writing weaker and not conventionally distinguishable from zero. Constructive-ratio and depth contrasts use conversation-level bootstrap resampling; adjacent-turn contrasts bootstrap conversations over assistant-to-user pairs.}\label{tab:key_contrast_significance}
\setlength{\tabcolsep}{5pt}
\renewcommand{\arraystretch}{1.10}
\resizebox{\textwidth}{!}{%
\begin{tabular}{llccc}
\toprule
Setting & Contrast & Effect & 95\% CI & p value \\
\midrule
WC coding & Scaffolded $-$ reference constructive ratio & +4.56 pp & [+4.15, +4.99] & $<.001$ \\
WC coding & Scaffolded $-$ reference post-answer depth & +2.23 turns & [+2.11, +2.35] & $<.001$ \\
LMSYS coding & Scaffolded $-$ reference constructive ratio & +2.50 pp & [+2.19, +2.83] & $<.001$ \\
LMSYS coding & Scaffolded $-$ reference post-answer depth & +1.42 turns & [+1.34, +1.52] & $<.001$ \\
SC coding & Scaffolded $-$ reference constructive ratio & +7.34 pp & [+5.29, +9.27] & $<.001$ \\
SC coding & Scaffolded $-$ reference post-answer depth & +3.36 turns & [+2.69, +4.01] & $<.001$ \\
WC writing & Scaffolded $-$ reference constructive ratio & +0.99 pp & [+0.82, +1.16] & $<.001$ \\
WC writing & Scaffolded $-$ reference post-answer depth & +3.13 turns & [+2.96, +3.30] & $<.001$ \\
LMSYS writing & Scaffolded $-$ reference constructive ratio & +1.09 pp & [+0.87, +1.31] & $<.001$ \\
LMSYS writing & Scaffolded $-$ reference post-answer depth & +2.10 turns & [+1.91, +2.28] & $<.001$ \\
SC writing & Scaffolded $-$ reference constructive ratio & +3.20 pp & [+1.55, +4.89] & $<.001$ \\
SC writing & Scaffolded $-$ reference post-answer depth & +3.38 turns & [+1.92, +4.64] & $<.001$ \\
WC coding & Adjacent scaffolded $-$ reference lift & +2.68 pp & [+2.22, +3.09] & $<.001$ \\
LMSYS coding & Adjacent scaffolded $-$ reference lift & +2.01 pp & [+1.54, +2.50] & $<.001$ \\
SC coding & Adjacent scaffolded $-$ reference lift & +6.21 pp & [+4.35, +8.10] & $<.001$ \\
WC writing & Adjacent scaffolded $-$ reference lift & +1.13 pp & [+0.88, +1.41] & $<.001$ \\
LMSYS writing & Adjacent scaffolded $-$ reference lift & +0.27 pp & [-0.01, +0.56] & .061 \\
SC writing & Adjacent scaffolded $-$ reference lift & +3.35 pp & [+0.80, +6.62] & .025 \\
\bottomrule
\end{tabular}%
}
\end{table*}

%% file: tables/table4.tex
\begin{table*}[t]
\footnotesize
\centering
\caption{
\textbf{Summary of key support--engagement associations across the three corpora.}
Constructive-ratio contrasts compare conversations with versus without scaffolded support using turn-weighted constructive user-turn ratios. Adjusted models report the association of scaffolded-support presence with constructive-turn count (Poisson count ratio) and with having at least one constructive turn (logit odds ratio). Adjacent-turn lift is the difference in the probability that the next user turn is constructive after scaffolded versus non-scaffolded reference assistant turns.
}
\label{tab:support_engagement_summary}

\setlength{\tabcolsep}{3pt}
\renewcommand{\arraystretch}{1.14}

\begin{tabular*}{\textwidth}{@{\extracolsep{\fill}}lrrrrrr@{}}
\toprule
Setting
& \makecell{Constructive ratio\\Has Scaf. / No Scaf.}
& \makecell{Difference\\(pp)}
& \makecell{Depth diff.\\(turns)}
& \makecell{Poisson\\count ratio}
& \makecell{Logit\\OR}
& \makecell{Adjacent lift\\(pp)} \\
\midrule
WildChat coding   & 0.101 / 0.056 & +4.6 & +2.23 & 1.852 & 1.761 & +2.68 \\
LMSYS coding      & 0.065 / 0.040 & +2.5 & +1.42 & 1.622 & 1.544 & +2.01 \\
ShareChat coding  & 0.163 / 0.090 & +7.3 & +3.36 & 1.893 & 2.140 & +6.21 \\
WildChat writing  & 0.027 / 0.017 & +1.0 & +3.13 & 1.569 & 1.437 & +1.13 \\
LMSYS writing     & 0.022 / 0.011 & +1.1 & +2.10 & 1.985 & 1.764 & +0.27 \\
ShareChat writing & 0.066 / 0.034 & +3.2 & +3.38 & 2.491 & 1.757 & +3.35 \\
\bottomrule
\end{tabular*}
\end{table*}

%% file: tables/table_fig3_offset_rate_sensitivity.tex
\begin{table*}[p]
\scriptsize
\centering
\caption{\textbf{Offset-rate sensitivity for scaffolded-support Poisson models.} The scaffolded-support rate ratio is statistically distinguishable from 1 in all six settings when user turns are represented as an exposure offset. The primary Fig.~\ref{fig:s2_conversation_association}b model uses raw constructive-turn counts with conversation length entered as a covariate; this sensitivity refits the same setting-specific mean model with \texttt{offset(log(user\_turns))}. P values test the two-sided null rate ratio of 1.}\label{tab:fig3_offset_rate_sensitivity}
\setlength{\tabcolsep}{5pt}
\renewcommand{\arraystretch}{1.08}
\resizebox{0.92\textwidth}{!}{%
\begin{tabular}{lccc}
\toprule
Setting & Offset rate ratio [95\% CI], p value & Pearson dispersion & Conversations \\
\midrule
WC coding & 1.57 [1.47, 1.68], $<.001$ & 1.29 & 31,878 \\
LMSYS coding & 1.47 [1.38, 1.56], $<.001$ & 1.13 & 32,114 \\
SC coding & 1.58 [1.31, 1.91], $<.001$ & 1.29 & 2,481 \\
WC writing & 1.36 [1.26, 1.46], $<.001$ & 1.15 & 39,534 \\
LMSYS writing & 1.70 [1.50, 1.94], $<.001$ & 1.12 & 21,023 \\
SC writing & 1.65 [1.27, 2.15], $<.001$ & 1.32 & 1,539 \\
\bottomrule
\end{tabular}%
}
\end{table*}

%% file: tables/table_section21_model_source_breakdown.tex
\begin{table*}[!htbp]
\scriptsize
\centering
\caption{\todayrevise{\textbf{Metadata breakdown for Section~\ref{subsec:2.1} descriptive engagement measures.} Ranges summarize available assistant-model or service-source levels with at least 100 conversations within each task setting. The full metadata rows, including passive, active, constructive, overall cognitive engagement, emotional expression and scaffolded-support percentages, are exported in the source CSV.}}\label{tab:section21_model_source_breakdown}
\setlength{\tabcolsep}{4pt}
\renewcommand{\arraystretch}{1.10}
\resizebox{0.94\textwidth}{!}{%
\begin{tabular}{lcccc}
\toprule
Setting & \todayrevise{Metadata levels} & Cognitive overall, \% & Constructive, \% & Scaffolded support, \% \\
\midrule
LMSYS coding & 23 & 31.55--49.70 (median 42.89) & 1.90--8.56 (median 5.34) & 4.80--43.87 (median 23.82) \\
LMSYS writing & 22 & 10.99--21.11 (median 14.19) & 0.54--2.77 (median 1.48) & 6.81--37.91 (median 18.68) \\
SC coding & 2 & 44.70--50.80 (median 47.75) & 6.57--15.43 (median 11.00) & 49.27--51.31 (median 50.29) \\
SC writing & 2 & 25.69--33.86 (median 29.77) & 1.19--5.41 (median 3.30) & 21.96--28.46 (median 25.21) \\
WC coding & 14 & 40.33--52.73 (median 51.07) & 6.66--13.70 (median 9.67) & 24.42--75.15 (median 64.94) \\
WC writing & 14 & 12.18--21.38 (median 17.69) & 0.90--3.30 (median 2.35) & 12.17--35.39 (median 25.85) \\
\bottomrule
\end{tabular}%
}
\end{table*}

%% file: tables/table_emotional_expression_summary.tex
\begin{table}[!htbp]
\footnotesize
\centering
\caption{\textbf{Explicit emotional-expression descriptor by task setting.} Emotional expression was coded in parallel with user-engagement depth and is reported descriptively; it is not part of the passive--active--constructive cognitive-depth hierarchy.}
\label{tab:emotional_expression_summary}
\setlength{\tabcolsep}{7pt}
\renewcommand{\arraystretch}{1.04}
\begin{tabular}{lrr}
\toprule
Setting & User turns & Emotional expression (\%) \\
\midrule
WildChat coding & 124,073 & 0.95 \\
WildChat writing & 163,705 & 0.57 \\
LMSYS coding & 107,065 & 0.91 \\
LMSYS writing & 77,906 & 0.21 \\
ShareChat coding & 11,541 & 3.38 \\
ShareChat writing & 7,395 & 0.34 \\
\midrule
Total / pooled & 491,685 & 0.74 \\
\bottomrule
\end{tabular}
\end{table}

%% file: tables/table_constructive_rate_context_logit.tex
\begin{table*}[p]
\scriptsize
\centering
\caption{\textbf{Context models for constructive engagement.} Both the any-constructive model and the constructive-rate sensitivity show the same organization: intentional framing, coding-oriented task ecology and longer exchanges are positively associated with constructive engagement. The rate model attenuates the length coefficients, indicating that part, but not all, of the depth gradient reflects greater opportunity to observe a constructive turn. Cells report odds ratios with 95\% confidence intervals and two-sided p values from conversation-robust standard errors. The reference category is unintentional framing, writing-oriented task, 2--3 user turns and WildChat.}
\label{tab:constructive_context_logit}
\setlength{\tabcolsep}{6pt}
\renewcommand{\arraystretch}{1.08}
\resizebox{0.88\textwidth}{!}{%
\begin{tabular}{lcc}
\toprule
Predictor & \makecell{Any constructive turn\\OR [95\% CI], p value} & \makecell{Constructive-turn rate\\OR [95\% CI], p value} \\
\midrule
Intentional framing & 3.03 [2.93, 3.13], $<.001$ & 2.78 [2.68, 2.88], $<.001$ \\
Coding task ecology & 1.89 [1.71, 2.08], $<.001$ & 1.66 [1.47, 1.88], $<.001$ \\
4--6 user turns & 2.07 [1.99, 2.15], $<.001$ & 1.13 [1.09, 1.18], $<.001$ \\
7+ user turns & 3.58 [3.42, 3.74], $<.001$ & 1.06 [1.01, 1.11], .010 \\
LMSYS Chat & 0.70 [0.68, 0.73], $<.001$ & 0.72 [0.69, 0.75], $<.001$ \\
ShareChat & 2.29 [2.12, 2.46], $<.001$ & 2.30 [2.13, 2.50], $<.001$ \\
\midrule
Conversations & 128,569 & 128,569 \\
User turns & -- & 491,685 \\
Constructive user turns & -- & 24,227 \\
\bottomrule
\end{tabular}%
}
\end{table*}

%% file: tables/table_section22_model_source_sensitivity.tex
\begin{table*}[p]
\scriptsize
\centering
\caption{\todayrevise{\textbf{Metadata sensitivity for Section~\ref{subsec:2.2} context models.} The dataset fixed-effect models match the context models reported for Section~\ref{subsec:2.2}. The metadata sensitivity replaces dataset fixed effects with available assistant-model or service-source indicators. Metadata levels with at least 100 conversations are represented as fixed-effect indicators; rarer levels are absorbed into the reference category. Cells report odds ratios with 95\% confidence intervals and two-sided p values from conversation-robust standard errors.}}\label{tab:section22_model_source_sensitivity}
\setlength{\tabcolsep}{5pt}
\renewcommand{\arraystretch}{1.10}
\resizebox{0.92\textwidth}{!}{%
\begin{tabular}{llcc}
\toprule
Outcome & Predictor & Dataset FE & \todayrevise{Metadata FE} \\
\midrule
Any constructive turn & Intentional framing & 3.03 [2.93, 3.13], $<.001$ & 3.05 [2.95, 3.16], $<.001$ \\
 & Coding task ecology & 1.89 [1.71, 2.08], $<.001$ & 1.89 [1.71, 2.08], $<.001$ \\
 & 4--6 user turns & 2.07 [1.99, 2.15], $<.001$ & 2.08 [2.00, 2.16], $<.001$ \\
 & 7+ user turns & 3.58 [3.42, 3.74], $<.001$ & 3.58 [3.42, 3.75], $<.001$ \\
\midrule
Constructive-turn rate & Intentional framing & 2.78 [2.68, 2.88], $<.001$ & 2.80 [2.70, 2.91], $<.001$ \\
 & Coding task ecology & 1.66 [1.47, 1.88], $<.001$ & 1.67 [1.48, 1.88], $<.001$ \\
 & 4--6 user turns & 1.13 [1.09, 1.18], $<.001$ & 1.14 [1.09, 1.18], $<.001$ \\
 & 7+ user turns & 1.06 [1.01, 1.11], .010 & 1.07 [1.02, 1.12], .008 \\
\midrule
Conversations & & 128,569 & 128,569 \\
\todayrevise{Metadata fixed effects} & & -- & 40 \\
\bottomrule
\end{tabular}%
}
\end{table*}

%% file: tables/table_section23_model_source_sensitivity.tex
\begin{table*}[p]
\scriptsize
\centering
\caption{\todayrevise{\textbf{Metadata sensitivity for Section~\ref{subsec:2.3} scaffolded-support models.} Pooled sensitivity models compare dataset fixed effects with available assistant-model or service-source fixed effects. Metadata levels with at least 100 conversations are represented as fixed-effect indicators; rarer levels are absorbed into the reference category. Rows report the scaffolded-support coefficient from models predicting constructive-turn counts, any constructive turn and constructive-turn rate. These checks supplement the main setting-level estimates in Fig.~\ref{fig:s2_conversation_association}b and Supplementary Table~\ref{tab:support_engagement_summary}.}}\label{tab:section23_model_source_sensitivity}
\setlength{\tabcolsep}{5pt}
\renewcommand{\arraystretch}{1.10}
\resizebox{0.86\textwidth}{!}{%
\begin{tabular}{lcc}
\toprule
Model & Dataset FE & \todayrevise{Metadata FE} \\
\midrule
Poisson count ratio & 2.38 [2.27, 2.48], $<.001$ & 2.35 [2.25, 2.45], $<.001$ \\
Logit odds ratio & 2.04 [1.96, 2.12], $<.001$ & 2.00 [1.92, 2.08], $<.001$ \\
Offset rate ratio & 1.94 [1.87, 2.01], $<.001$ & 1.91 [1.85, 1.98], $<.001$ \\
\midrule
Conversations & 128,569 & 128,569 \\
\todayrevise{Metadata fixed effects} & -- & 40 \\
\bottomrule
\end{tabular}%
}
\end{table*}

%% file: tables/table_section24_support_form_model_source_sensitivity.tex
\begin{table*}[!htbp]
\scriptsize
\centering
\caption{\todayrevise{\textbf{Metadata sensitivity for Section~\ref{subsec:2.4} support-form models.} The sensitivity is restricted to conversations containing scaffolded support and enters the six non-exclusive support-form indicators jointly. Cells report odds ratios for constructive-turn rate with 95\% confidence intervals and two-sided p values from conversation-robust standard errors. This metadata check complements, but does not replace, the descriptive percentage-point support-form contrasts in Fig.~\ref{fig:support_form_supply}.}}\label{tab:section24_support_form_model_source_sensitivity}
\setlength{\tabcolsep}{5pt}
\renewcommand{\arraystretch}{1.10}
\resizebox{0.82\textwidth}{!}{%
\begin{tabular}{lcc}
\toprule
Support form & Dataset FE & \todayrevise{Metadata FE} \\
\midrule
M1 feedback & 2.00 [1.90, 2.10], $<.001$ & 2.02 [1.92, 2.13], $<.001$ \\
M2 hinting & 1.37 [1.30, 1.44], $<.001$ & 1.36 [1.29, 1.43], $<.001$ \\
M3 instructing & 0.92 [0.88, 0.97], $<.001$ & 0.92 [0.88, 0.97], $<.001$ \\
M4 explaining & 2.81 [2.64, 2.99], $<.001$ & 2.77 [2.60, 2.94], $<.001$ \\
M5 modelling & 0.77 [0.74, 0.81], $<.001$ & 0.78 [0.74, 0.82], $<.001$ \\
M6 questioning & 0.69 [0.64, 0.73], $<.001$ & 0.70 [0.65, 0.75], $<.001$ \\
\midrule
Scaffolded conversations & 66,816 & 66,816 \\
\todayrevise{Metadata fixed effects} & -- & 39 \\
\bottomrule
\end{tabular}%
}
\end{table*}

%% file: tables/table_integrated_regression.tex
\begin{table*}[p]
\scriptsize
\centering
\caption{\textbf{Integrated adjacent-turn regression combining user state, assistant scaffolding and model controls.} Prior user state is the largest predictor of next-turn constructive engagement, broad scaffolded support remains positive after controls and M4 explaining is the most stable positive support-form coefficient. The scaffolded-support row is estimated in a broad scaffolded-support model; support-form rows are estimated in a decomposed support-form model with broad scaffolded-support presence and co-occurring M1--M6 support forms; \todayrevise{all six support-form labels are shown to avoid restricting the table to selected focal forms.} Estimates are odds ratios with 95\% confidence intervals and two-sided p values from conversation-cluster robust standard errors. \todayrevise{The metadata-block LR row compares the pooled dataset-fixed-effect and metadata-fixed-effect adjacent-turn specifications: the metadata design contains 43 non-reference assistant-model or service-source indicators in place of two dataset indicators, giving 41 net degrees of freedom; intercepts and reference levels are not counted.}}\label{tab:integrated_regression}
\setlength{\tabcolsep}{4pt}
\renewcommand{\arraystretch}{1.10}
\resizebox{\textwidth}{!}{%
\begin{tabular}{lccc}
\toprule
Predictor & Pooled dataset FE & \todayrevise{Pooled metadata FE} & WildChat model FE \\
\midrule
Scaffolded support & 1.50 [1.44, 1.55], $<.001$ & 1.47 [1.41, 1.52], $<.001$ & 1.56 [1.49, 1.63], $<.001$ \\
Prior user constructive & 8.90 [8.45, 9.38], $<.001$ & 8.55 [8.11, 9.01], $<.001$ & 9.11 [8.53, 9.74], $<.001$ \\
Prior user active & 2.62 [2.52, 2.72], $<.001$ & 2.57 [2.47, 2.67], $<.001$ & 2.85 [2.71, 3.00], $<.001$ \\
Prior user passive & 2.08 [1.76, 2.44], $<.001$ & 2.07 [1.76, 2.44], $<.001$ & 2.32 [1.85, 2.90], $<.001$ \\
Intentional framing & 1.40 [1.35, 1.46], $<.001$ & 1.42 [1.37, 1.48], $<.001$ & 1.40 [1.33, 1.47], $<.001$ \\
Coding task & 1.37 [1.22, 1.54], $<.001$ & 1.37 [1.22, 1.54], $<.001$ & 1.21 [1.05, 1.40], .009 \\
M1 feedback & 0.79 [0.73, 0.86], $<.001$ & 0.80 [0.74, 0.86], $<.001$ & 0.74 [0.67, 0.82], $<.001$ \\
M2 hinting & 1.02 [0.95, 1.10], .546 & 1.03 [0.96, 1.11], .448 & 1.07 [0.98, 1.16], .122 \\
M3 instructing & 0.91 [0.86, 0.97], .005 & 0.93 [0.87, 0.99], .020 & 0.94 [0.87, 1.01], .099 \\
M4 explaining & 2.15 [1.99, 2.32], $<.001$ & 2.12 [1.96, 2.30], $<.001$ & 2.11 [1.92, 2.31], $<.001$ \\
M5 modelling & 0.78 [0.73, 0.83], $<.001$ & 0.78 [0.73, 0.84], $<.001$ & 0.80 [0.73, 0.86], $<.001$ \\
M6 questioning & 0.82 [0.72, 0.93], .002 & 0.84 [0.74, 0.96], .008 & 1.08 [0.93, 1.25], .340 \\
\midrule
\todayrevise{Metadata block} & Reference & LR $\chi^2_{41}=714.4$, $p<.001$ & LR $\chi^2_{13}=553.2$, $p<.001$ \\
\bottomrule
\end{tabular}%
}
\end{table*}

%% file: tables/table_integrated_scaffolding_block_tests.tex
\begin{table*}[p]
\scriptsize
\centering
\caption{\textbf{Incremental contribution of scaffolding features in integrated adjacent-turn models.} Broad scaffolded support, support-form descriptors and the full scaffolding block each improve model fit after user/context controls, including when dataset fixed effects are replaced by \todayrevise{available assistant-model or service-source fixed effects}. Likelihood-ratio block tests compare nested logistic regressions for whether the next user turn is constructive.}\label{tab:integrated_scaffolding_block_tests}
\setlength{\tabcolsep}{4pt}
\renewcommand{\arraystretch}{1.08}
\resizebox{\textwidth}{!}{%
\begin{tabular}{llcc}
\toprule
Scope & Added feature block & LR $\chi^2$ (df) & p value \\
\midrule
six task settings, dataset FE & Broad scaffolded support after user/context controls & 584.6 (1) & $<.001$ \\
six task settings, dataset FE & M1--M6 descriptors within scaffolded support & 896.7 (6) & $<.001$ \\
six task settings, dataset FE & Full scaffolding block after user/context controls & 1481.3 (7) & $<.001$ \\
six task settings, \todayrevise{metadata FE} & Broad scaffolded support after user/context controls & 509.7 (1) & $<.001$ \\
six task settings, \todayrevise{metadata FE} & M1--M6 descriptors within scaffolded support & 829.8 (6) & $<.001$ \\
six task settings, \todayrevise{metadata FE} & Full scaffolding block after user/context controls & 1339.6 (7) & $<.001$ \\
WildChat only, model FE & Broad scaffolded support after user/context controls & 418.6 (1) & $<.001$ \\
WildChat only, model FE & M1--M6 descriptors within scaffolded support & 519.4 (6) & $<.001$ \\
WildChat only, model FE & Full scaffolding block after user/context controls & 937.9 (7) & $<.001$ \\
\bottomrule
\end{tabular}%
}
\end{table*}

%% file: tables/table_prior_state_support_form_interactions.tex
\begin{table*}[p]
\scriptsize
\centering
\caption{\textbf{Prior user state and support form jointly structure adjacent-turn constructive engagement.} Prior-state $\times$ support-form interactions improve model fit, and M4 explaining is positive after constructive, active and passive prior user states; M1 feedback is more state-dependent. The first panel reports likelihood-ratio tests; the second panel reports \todayrevise{pooled metadata-adjusted odds ratios} with 95\% confidence intervals and two-sided p values.}\label{tab:prior_state_support_form_interactions}
\setlength{\tabcolsep}{4pt}
\renewcommand{\arraystretch}{1.08}
\begin{tabularx}{\textwidth}{>{\raggedright\arraybackslash}p{0.20\textwidth}>{\raggedright\arraybackslash}p{0.18\textwidth}>{\raggedright\arraybackslash}p{0.22\textwidth}>{\raggedright\arraybackslash}X}
\toprule
Panel & Scope / prior state & Term or test & Estimate \\
\midrule
Interaction block & six task settings, dataset FE & Prior state $\times$ M1--M6 & LR $\chi^2_{18}=305.4$, $<.001$ \\
\todayrevise{Interaction block} & six task settings, \todayrevise{metadata FE} & Prior state $\times$ M1--M6 & LR $\chi^2_{18}=297.3$, $<.001$ \\
Interaction block & WildChat only, model FE & Prior state $\times$ M1--M6 & LR $\chi^2_{18}=186.0$, $<.001$ \\
\midrule
\todayrevise{State-stratified metadata FE} & prior constructive & M1 feedback & 0.72 [0.64, 0.81], $<.001$ \\
 & prior constructive & M2 hinting & 0.91 [0.79, 1.04], .155 \\
 & prior constructive & M3 instructing & 0.92 [0.81, 1.05], .236 \\
 & prior constructive & M4 explaining & 1.72 [1.45, 2.04], $<.001$ \\
 & prior constructive & M5 modelling & 0.81 [0.70, 0.93], .003 \\
 & prior constructive & M6 questioning & 0.95 [0.72, 1.25], .694 \\
 & prior active & M1 feedback & 0.83 [0.72, 0.95], .008 \\
 & prior active & M2 hinting & 1.14 [1.03, 1.27], .016 \\
 & prior active & M3 instructing & 0.98 [0.89, 1.08], .722 \\
 & prior active & M4 explaining & 2.13 [1.88, 2.41], $<.001$ \\
 & prior active & M5 modelling & 0.83 [0.76, 0.91], $<.001$ \\
 & prior active & M6 questioning & 0.75 [0.61, 0.93], .009 \\
 & prior passive & M1 feedback & 2.16 [1.16, 4.04], .015 \\
 & prior passive & M2 hinting & 1.79 [0.95, 3.37], .071 \\
 & prior passive & M3 instructing & 1.52 [0.67, 3.43], .312 \\
 & prior passive & M4 explaining & 2.88 [1.55, 5.35], $<.001$ \\
 & prior passive & M5 modelling & 0.46 [0.20, 1.04], .061 \\
 & prior passive & M6 questioning & 2.19 [1.12, 4.27], .021 \\
\bottomrule
\end{tabularx}
\end{table*}